\PassOptionsToPackage{unicode}{hyperref}
\PassOptionsToPackage{hyphens}{url}
\documentclass[
  12pt,
  a4paper,
]{article}
\usepackage{amsmath,amssymb}
\usepackage{lmodern}
\usepackage{setspace}
\usepackage{authblk}
\usepackage{iftex}
\ifPDFTeX
  \usepackage[T1]{fontenc}
  \usepackage[utf8]{inputenc}
  \usepackage{textcomp} 
\else 
  \usepackage{unicode-math}
  \defaultfontfeatures{Scale=MatchLowercase}
  \defaultfontfeatures[\rmfamily]{Ligatures=TeX,Scale=1}
\fi
\IfFileExists{upquote.sty}{\usepackage{upquote}}{}
\IfFileExists{microtype.sty}{
  \usepackage[]{microtype}
  \UseMicrotypeSet[protrusion]{basicmath} 
}{}
\makeatletter
\@ifundefined{KOMAClassName}{
  \IfFileExists{parskip.sty}{%
    \usepackage{parskip}
  }{
    \setlength{\parindent}{0pt}
    \setlength{\parskip}{6pt plus 2pt minus 1pt}}
}{
  \KOMAoptions{parskip=half}}
\makeatother
\usepackage{xcolor}
\usepackage[left=2.54cm, right=2.54cm, top=2.54cm, bottom=2.54cm]{geometry}
\usepackage{longtable,booktabs,array}
\usepackage{calc} 
\usepackage{etoolbox}
\makeatletter
\patchcmd\longtable{\par}{\if@noskipsec\mbox{}\fi\par}{}{}
\makeatother
\IfFileExists{footnotehyper.sty}{\usepackage{footnotehyper}}{\usepackage{footnote}}
\makesavenoteenv{longtable}
\usepackage{graphicx}
\makeatletter
\def\maxwidth{\ifdim\Gin@nat@width>\linewidth\linewidth\else\Gin@nat@width\fi}
\def\maxheight{\ifdim\Gin@nat@height>\textheight\textheight\else\Gin@nat@height\fi}
\makeatother
\setkeys{Gin}{width=\maxwidth,height=\maxheight,keepaspectratio}
\makeatletter
\def\fps@figure{htbp}
\makeatother
\newlength{\cslhangindent}
\newlength{\csllabelwidth}
\newlength{\cslentryspacingunit} 
\newenvironment{CSLReferences}[2] 
 {
  \setlength{\parindent}{0pt}
  \ifodd #1
  \let\oldpar\par
  \def\par{\hangindent=\cslhangindent\oldpar}
  \fi
  \setlength{\parskip}{#2\cslentryspacingunit}
 }%
 {}
\usepackage{calc}

\usepackage{float}
\usepackage{caption}
\usepackage{accents}
\usepackage{amsthm}
\newtheorem*{remarkun}{Remark}
\newtheorem*{propun}{Proposition}
\usepackage{booktabs}
\usepackage{longtable}
\usepackage{array}
\usepackage{multirow}
\usepackage{wrapfig}
\usepackage{float}
\usepackage{colortbl}
\usepackage{pdflscape}
\usepackage{tabu}
\usepackage{threeparttable}
\usepackage{threeparttablex}
\usepackage[normalem]{ulem}
\usepackage{makecell}
\usepackage{xcolor}
\ifLuaTeX
  \usepackage{selnolig}  
\fi
\IfFileExists{bookmark.sty}{\usepackage{bookmark}}{\usepackage{hyperref}}
\IfFileExists{xurl.sty}{\usepackage{xurl}}{} 
\hypersetup{
  pdftitle={Logistic Regression Equivalence: A Framework for Comparing Logistic Regression Models Across Populations},
  hidelinks,
  pdfcreator={LaTeX via pandoc}}

\title{Logistic Regression Equivalence: A Framework for Comparing Logistic Regression Models Across Populations}
\author[1,2]{Guy Ashiri-Prossner}
\author[1]{Yuval Benjamini}
\affil[1]{\footnotesize Department of Statistics and Data Science, The Hebrew University of Jerusalem}
\affil[2]{\footnotesize National Institute for Testing \& Evaluation}
\date{February 28, 2022}

\usepackage{amsthm}

\newtheorem{proposition}{Proposition}[section]

\theoremstyle{definition}

\theoremstyle{definition}

\theoremstyle{definition}

\theoremstyle{definition}

\theoremstyle{remark}

\begin{document}
\maketitle

\setstretch{1.5}
\hypertarget{abstract}{%
\section*{Abstract ~}\label{abstract}}
\addcontentsline{toc}{section}{Abstract ~}

In this paper we discuss how to evaluate the differences between fitted logistic regression models across sub-populations.
Our motivating example is in studying computerized diagnosis for learning disabilities, where sub-populations based on gender may or may not require separate models.
In this context, significance tests for hypotheses of no difference between populations may provide perverse incentives, as larger variances and smaller samples increase the probability of not-rejecting the null.
We argue that equivalence testing for a prespecified tolerance level on population differences incentivizes accuracy in the inference.
We develop a cascading set of equivalence tests, in which each test addresses a different aspect of the model: the way the phenomenon is coded in the regression coefficients, the individual predictions in the per example log odds ratio and the overall accuracy in the mean square prediction error.
For each equivalence test, we propose a strategy for setting the equivalence thresholds.
The large-sample approximations are validated using simulations. For diagnosis data, we show examples for equivalent and non-equivalent models.

\hypertarget{introduction}{%
\section{Introduction}\label{introduction}}

A common challenge in developing psychometric tools is deciding when to design separate models or tests for specific sub-populations or demographic groups. Responses to tests often vary along demographic covariates such as sex, age, mother tongue and socio-economical status (Byrne (\protect\hyperlink{ref-byrne1988measuring}{1988}), Collins and Gleaves (\protect\hyperlink{ref-collins1998race}{1998})). These variations usually persist even after conditioning on the concept to be measured (\protect\hyperlink{ref-byrne2014factorial}{Byrne and van de Vijver 2014}). The population response distribution may also change over time, potentially requiring model recalibration. Ideally, for a given ability level, the distribution of test scores should be identical across sub-populations. However, from an organizational standpoint, introducing and validating tests for any potential sub-population may not be feasible: beyond the resources required to develop, calibrate and administer multiple exams, there also costs in having experts trained to interpret the results from multiple models. In high-stakes exams, the question of model calibration for sub-populations can affect access to higher education or to jobs in the public sector. It is therefore important to identify transparent metrics and criteria for similarity between populations in the context of such examinations.

In this paper we focus on diagnostic tests with binary outcomes. We are motivated by automated exams for detecting learning disabilities in Israeli higher education. Such diagnostic exams are constructed in a supervised manner, trying to match an annotation considered to be the ground truth, for example one based on expert assessment. Due to its popularity, we restrict the analysis to a logistic regression-based classifier (\protect\hyperlink{ref-cramer2002origins}{Cramer 2002}). For logistic regression (simply as a classifier, unrelated to the popular DIF detection method by Swaminathan and Rogers (\protect\hyperlink{ref-swaminathan1990detecting}{1990})), a simple way to measure invariance is by including interaction terms for the inputs to the regression, and comparing the resulting models (\protect\hyperlink{ref-hosmer2013applied}{Hosmer Jr et al. 2013}).

The influence of demographic covariates on diagnostic outcomes has been studied extensively in educational statistics. Under item response theory, Steinberg and Thissen (\protect\hyperlink{ref-steinberg2006using}{2006}) developed methods for detecting bias in individual items using effect size, with respect to sub-populations. Given an item and its responses data for two (or more) distinct populations, these methods compare (either graphically or by computation) the differences between the two estimated item functions. Many other differential item functioning (DIF) detection methods are available, see Özdemir (\protect\hyperlink{ref-ozdemir2015comparison}{2015}), Holland and Wainer (\protect\hyperlink{ref-holland2012differential}{2012}), Magis and De Boeck (\protect\hyperlink{ref-magis2011identification}{2011}) and Martinková et al. (\protect\hyperlink{ref-martinkova2017checking}{2017}) for a review. More generally, the \emph{measurement invariance} (\protect\hyperlink{ref-meredith1993measurement}{Meredith 1993}) framework is concerned with a systematic study of the multivariate distributions of latent factors derived from test-scores. Measurement invariance requires that the association between items or test-scores and the latent variables derived from them would not depend on demographic covariates (\protect\hyperlink{ref-van2015measurement}{van de Schoot et al. 2015}). Because independence can break at various points along the analysis, the assessment of measurement invariance follows a cascade of statistical tests (Putnick and Bornstein (\protect\hyperlink{ref-putnick2016measurement}{2016}), Vandenberg and Lance (\protect\hyperlink{ref-vandenberg2000review}{2000})) sensitive to find differences in distribution of the test-scores, the loadings of the factors, biases in the factor distributions and more.

Consider a prediction model trained on data from sub-population \(A\) and a distinct dataset from sub-population \(B\). There are two major related problems with using significance tests for deciding whether to use the model `as is', re-fit the model (by adding data from sub-population \(B\)) or fit a separate model. The first is that the null hypothesis is probably never exactly correct: for any meaningful partitioning of a population into sub-populations, there will be some difference, perhaps small, in the distribution of test scores given the categorical classification. This is known as the null hypothesis fallacy (\protect\hyperlink{ref-brenner1985evidence}{Brenner 1985}). As sample-sizes increase, invariance tests become powerful to detect even such minor differences. Examining recent developments in the measurement invariance literature we see that the criterion is usually relaxed by using Bayesian methods for estimating the biases (\protect\hyperlink{ref-verhagen2013bayesian}{Verhagen and Fox 2013}) or by using indifference regions. The second problem is that often the test-developers incur costs by rejecting invariance. Under the significance testing framework, their incentive may hypothetically be to use a smaller sample size or less precise measurements so that the null hypothesis would not be rejected.

We propose addressing these two challenges with equivalence testing. Equivalence tests are designed to ascertain that the difference between sub-populations does not exceed a predetermined acceptable threshold. There are several recent works in psychometrics using equivalence testing for comparing models across groups. Casabianca and Lewis (\protect\hyperlink{ref-casabianca2018statistical}{2018}) offer an equivalence test for the Mantel-Haenszel test of differential item functioning (\protect\hyperlink{ref-holland1988differential}{Holland and Thayer 1988}); Weigold et al. (\protect\hyperlink{ref-weigold2016equivalence}{2016}) use equivalence testing for comparing the results of paper-and-pencil surveys against computer-administered surveys. Outside of psychometrics, equivalence tests have been used to compare coefficients in linear or generalized linear regression models (Counsell and Cribbie (\protect\hyperlink{ref-counsell2015equivalence}{2015}), Jonkman and Sidik (\protect\hyperlink{ref-jonkman2009equivalence}{2009})). Equivalence tests for individual sample predictions have been proposed for the predicted expected value (W. Liu et al. (\protect\hyperlink{ref-liu2009assessing}{2009}) and Wei Liu (\protect\hyperlink{ref-liu2010simultaneous}{2010})), log odds (\protect\hyperlink{ref-siqueira2008active}{Siqueira et al. 2008}) and probability in binary regressions (Stevens and Anderson-Cook (\protect\hyperlink{ref-stevens2017comparing}{2017a}), Stevens and Anderson-Cook (\protect\hyperlink{ref-stevens2017quantifying}{2017b})). Most similar to our approach, Dette et al. (\protect\hyperlink{ref-dette2018equivalence}{2018}) proposes bootstrap-based equivalence test for comparing nonlinear regression models in terms of the \(L^1\) of \(L^2\) distance between predictions.

In this paper, we develop a more comprehensive approach for comparing logistic regression diagnostic tests across sub-populations. We identify three distinct stages of equivalence:

\begin{enumerate}
\def\labelenumi{\arabic{enumi}.}
\item
  \textbf{Descriptive equivalence} of models is achieved when two models describe the relation between predictors and outcome in a similar manner. We will check for descriptive equivalence by comparing the regression coefficient vectors. Descriptive equivalence (DE) implies the effect of the corresponding coefficients is sufficiently similar.
\item
  \textbf{Individual predictive equivalence} of models is achieved when two models yield the similar predictions for a set of observations. We will check for this equivalence by comparing the log-odds produced by the models on a new set of examples. Individual predictive equivalence (IPE) implies the outputted predictions remain stable, within a predefined threshold, even if the models are exchanged.
\item
  \textbf{Performance equivalence} of models is achieved when the prediction accuracy of the two models is similar. We will check for this equivalence by comparing the Brier scores. Performance equivalence (PE) implies that the prediction accuracy would not be hurt if the models are exchanged.
\end{enumerate}

The three equivalences capture different aspects in the development of predictive models.
The choice of equivalence method depends on the particular needs of a researcher or an organization: Should they wish to test whether the two coefficient vectors are similar, descriptive equivalence would fit. In case they want to test whether the models produce similar predictions for a specific test set, individual predictive equivalence would fit (even if descriptive equivalence is not achieved or tested for). If they want to test only for similar overall accuracy of prediction, performance equivalence would fit (regardless of the two other methods).

We prove that at appropriate thresholds these equivalences form a cascade, meaning that descriptive equivalence implies individual predictive equivalence. Similarly, individual predictive equivalence implies performance equivalence. In practice, the different equivalences correspond to different usages of the models, and we therefore expect that equivalence bounds would be determined separately for each stage. We propose methods for choosing the equivalence bounds for each stage of the cascade.

The rest of the paper is organized as follows. The remainder of the section will be devoted to a brief introduction of equivalence testing. In Section \ref{methods} we derive equivalence tests for the three stages in our cascade. For each stage we discuss how the threshold should be parameterized. Section \ref{sim-stud} shows in simulations that the equivalence methods work as stated. In Section \ref{matal}, we study the effects of sex on the diagnostic test learning disabilities used by the Israel Higher Education Council. The data analysed in this section was re-randomized.

\hypertarget{equivalence-testing}{%
\subsection{Equivalence Testing}\label{equivalence-testing}}

\emph{Equivalence testing} (see Wellek (\protect\hyperlink{ref-wellek2010testing}{2010}) for an introduction) develops statistical test (e.g.~for the difference between two groups) to demonstrate with high probability that an effect size is negligible. Equivalence testing provides a conceptual change compared to usual significance testing, as it moves the burden of proof (\protect\hyperlink{ref-dolado2014equivalence}{Dolado et al. 2014}). Instead of assuming no effect and proving that an effect exists, the test inverts the direction of the hypotheses.
Acknowledging that very small deviations from the null cannot be disproven, equivalence testing requires the researcher to decide on a maximal acceptable difference between the populations. Then, the null hypothesis will be that the effect size is at least \(\epsilon\). It is up to the researcher to prove the alternative that the effect is smaller.

For example, consider samples of size \(n\) from two populations: \(X^A_i\sim\mathcal{N}(\mu^A,1)\), \(X^B_i\sim\mathcal{N}(\mu^B,1)\) and denote the difference of expectations \(\mu=\mu^A-\mu^B\).
If the goal of the researcher is to prove that expectations are not equal, the appropriate hypotheses for significance testing are \(H_0: \mu=0,~ H_1: \mu\ne0\) and the test is of the form \(\left\{\left|\bar{X^A}-\bar{X}^B\right|>z_{1-\alpha/2}\cdot \sqrt{2/n}\right\}\). On the other hand, if the goal us to prove that the effect size is smaller than \(\epsilon\), the equivalence testing hypotheses are \(H_0: |\mu|\geq\epsilon,~ H_1: |\mu|<\epsilon\) and the test is of the form (Wellek (\protect\hyperlink{ref-wellek2010testing}{2010}), Chapter 4):

\begin{equation}
\left\{\left|\bar{x}_n\right|<\sqrt{\chi^2_{1,\alpha}(n\cdot\epsilon^2/2)}/\sqrt{n/2}\right\}.
\label{eq:equiv-chi}
\end{equation}

As Robinson et al. (\protect\hyperlink{ref-robinson2005regression}{2005}) observes, the burden of proof for determining equivalence shifts to the scientist to reject this hypothesis, or in other words, the scientist needs to show the difference between models is small enough to be acceptable. By setting this null hypothesis, the parameter-region of equivalence grows, rather than narrows, as sample size increases. In standard (null-effect) hypothesis testing we assume the population means (for example) to be equal. Then, we use the data to disprove our null hypothesis of no difference. In equivalence testing we assume that the population means differ (by at least \(\epsilon\)). Then, we use the data to prove equality. Moreover, equivalence testing changes the way a study should be designed, as explained by Walker and Nowacki (\protect\hyperlink{ref-walker2011understanding}{2011}): The need to determine the equivalence threshold \(\epsilon\) \emph{before} any data is collected might be the most important change. By inverting the hypotheses direction, equivalence testing also bounds the probability of mistakenly finding the populations to be equivalent (\protect\hyperlink{ref-barker2002assessing}{Barker et al. 2002}).

\hypertarget{methods}{%
\section{Methods}\label{methods}}

The goal of this work is to provide a framework for comprehensive comparison of two logistic regression models, overcoming common problems. We propose a cascading set of equivalence tests, inspired by the measurement invariance framework. The equivalence tests we develop for the coefficient vectors and the log-odds-ratio in the logistic regression are based on their asymptotic distributions (see Peng et al. (\protect\hyperlink{ref-peng2002introduction}{2002}) for an introduction).

\hypertarget{DSN}{%
\subsection{Data Structure and Notations}\label{DSN}}

For a sample of \(n\) individuals, let \(\left(X_{1},...,X_{p-1}\right)_{i}\in\mathbb{R}^{p-1}\) be the vector of covariates for individual \(i\), \(X_{0i}=1\) the intercept and \(Y_i\in\{0,1\}\) the binary response of interest, for \(i=1,...,n\). The logistic regression model is defined as

\[E[Y_i|X_i=x_i]=P(Y_i=1|X_i=x_i)=\frac{e^{\beta^Tx_i}}{1+e^{\beta^Tx_i}}.\]
For a given \(\hat{\beta}\) we denote the linear predictor by \(\hat{\theta}_i=\hat{\beta}^Tx_i\), and the predicted probability by \(\hat{\pi}_i= \frac{e^{\hat{\theta}_i}}{1+e^{\hat{\theta}_i}}\).

We assume that our data consists of samples from two sub-populations, denoted by \(A\) and \(B\).
The samples of population \(A\) are split into distinct \emph{train} and \emph{test} sets (In practice, these sets may be collected separately). A logistic regression model \(M^A\) is fit using the training data (\(X_A^{train},y_A^{train}\)). The model can be described by its coefficient vector \(\hat{\beta}^A\) and its coefficients' covariance matrix \(V^A\). The data of a population \(B\) (\(X_B,y_B\)) will yield model \(M^B\) that can be described by \(\hat{\beta}^B, V^B\).

For comparing the predictions, we look to the vectors of linear predictors and binary predictions that are associated with two different coefficient vectors and a single test population. We use the superscript to denote the coefficient vector, and the subscript to denote the test population. Hence, the vector of linear predictors outputted by using \(M^A\) on the \emph{test} set of population \(A\) (\(X_A^{test}\)) is denoted \(\hat{\theta}_A^A\), the vector of output probabilities is denoted \(\hat{\pi}_A^A\) and the vector of binary predictions is denoted \(\hat{y}_A^A\). By using \(M^B\) on the same \emph{test} set of population \(A\) (\(X_A^{test}\)) we will obtain \(\hat{\theta}_A^B, \hat{\pi}_A^B, \hat{y}_A^B\). Although we do not use \(\hat{y}\) in our methods, it is notes here for the sake of completeness.

\begin{figure}[!h]

{\centering \includegraphics[width=0.8\linewidth]{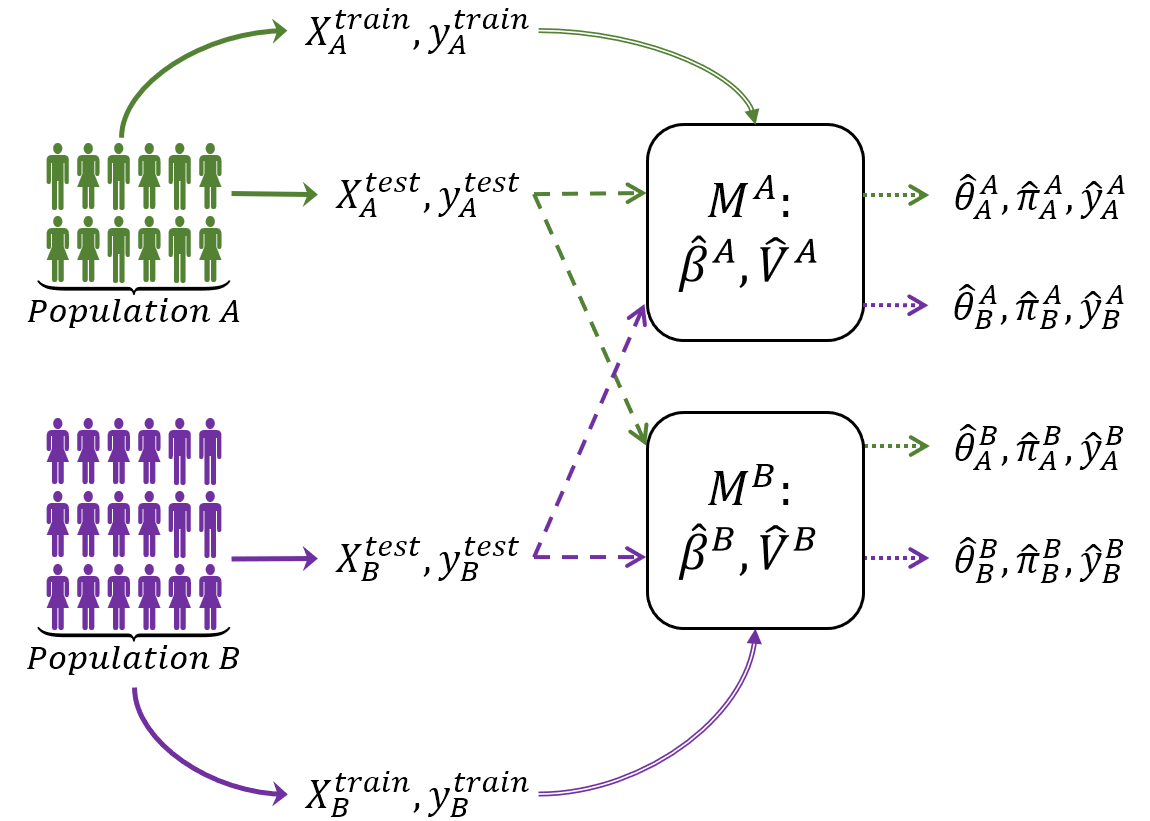} 

}

\caption{Data structure. Solid lines indicate splitting of population data; double lines indicate data used for training a model; dashed lines indicate data used as model input; dotted lines indicate model outputs. Subscript indicated population and superscript indicates model.}\label{fig:pops1}
\end{figure}

We suggest the following comparisons:

\begin{enumerate}
\def\labelenumi{\arabic{enumi}.}
\item
  Descriptive equivalence - Compare the coefficient vector estimates \(\hat{\beta}^A, \hat{\beta}^B\) in order to compare the effects of the different covariates on the predictor \(\hat{\theta}\). The test accounts for the sampling distribution of the training sets, characterized by the covariances \(V^A\) and \(V^B\).
\item
  Individual predictive equivalence - Compare the linear predictors of two trained models on a single test population. \(\hat{\theta}^A_A, \hat{\theta}^B_A\). The test accounts for the sampling distribution of the test sets, conditional on the estimated models.
\item
  Performance equivalence - Compare the overall prediction accuracy \((y_i-\hat{\pi}_i)\) of the two models for a single test population. The test accounts for the sampling distribution of the test sets, conditional on the estimated models.
\end{enumerate}

The three tests offer different perspectives of model equivalence. Much like in the measurement equivalence framework, they do follow a cascading order:

\begin{proposition}
\protect\hypertarget{prp:propo1}{}\label{prp:propo1}Descriptive equivalence implies individual predictive equivalence, using a proper equivalence threshold.

\normalfont More specifically, descriptive equivalence of models with equivalence threshold \(\epsilon_\beta\) implies individual predictive equivalence over test set \(X\) of size \(m\) with equivalence threshold \(\epsilon_\theta=\epsilon_\beta\sqrt{\lambda_1}\sqrt{{\mu_X}^T\mu_X+tr(\Sigma_X)}\), where \(\mu_X\) is the expected value vector of the test set covariates and \(\Sigma_X\) is its covariance matrix.
\end{proposition}

\begin{proposition}
\protect\hypertarget{prp:propo2}{}\label{prp:propo2}Individual predictive equivalence implies performance equivalence, using a proper equivalence threshold.

\normalfont More specifically, individual predictive equivalence of models over test set \(X\) of size \(m\) wirh equivalence threshold \(\epsilon_\theta\) implies performance equivalence over the same test set with equivalence threshold \(\epsilon_B=\exp(2\epsilon_\theta)\).
\end{proposition}

The proofs appear in Section \ref{cascade}.

We will use equivalence tests in order to provide robust performance with large sample sizes. In addition to specifying the required significance level \(\alpha\), equivalence testing requires specifying an equivalence threshold \(\epsilon\). As can be seen in \eqref{eq:equiv-chi}, the \(\epsilon\) value should be selected in terms of the test statistic distribution rather than the data itself, which might not be intuitive to the practitioner. We therefore identify for each test a \emph{sensitivity level} parameter \(\delta\) that can be set externally, and a conversion function \(\epsilon=f(\delta)\) to set the equivalence threshold for the test. Each of the suggested methods is accompanied by a suggestion for choosing a proper \(\delta\) value (which in turn yields an equivalence threshold \(\epsilon\)).

\hypertarget{comparing-beta-hat}{%
\subsection{Descriptive Equivalence: Testing Regression Coefficients}\label{comparing-beta-hat}}

In the logistic regression model, the regression coefficients code the effects of each covariate on the outcome. The coefficient vectors are therefore the strongest indication that the prediction models describe the relation of the response to the covariates in a similar manner. Let \(\beta^A\) and \(\beta^B\) be the coefficient vectors obtained from logistic regression models on populations \(A\) and \(B\). Denote by \(q = \beta^A - \beta^B\) the difference between these vectors. We say models \(M_A\) and \(M_B\) obtain descriptive equivalence if the size of \(q\) is sufficiently small. We describe here a chi-square-based equivalence-test, sensitive to detect dense differences between the coefficient vectors. It is not hard to construct a similar test (or test-family) that is sensitive to differences in an individual coordinate; see Wells et al. (\protect\hyperlink{ref-wells2009range}{2009}) and Casabianca and Lewis (\protect\hyperlink{ref-casabianca2018statistical}{2018}).

Our test examines the difference between the logistic regression coefficient vectors that were estimated on samples from the two populations. In large samples, the maximum likelihood estimator (MLE) for \(\beta\) is approximately normally distributed around the true parameter \(\hat{\beta}\dot{\sim}\mathcal{N}(\beta,V)\). See Hosmer Jr et al. (\protect\hyperlink{ref-hosmer2013applied}{2013}) for further details. The estimated vectors \(\hat{\beta}^A\) and \(\hat{\beta}^B\) are asymptotically normally distributed. Therefore, the observed difference vector \(\hat{q} = \hat{\beta}^A - \hat{\beta}^B\) is also normal, \(\hat{q}\stackrel{\cdot}{\sim}\mathcal{N}(q, V^q)\). The covariance matrix \(V^q\) is the sum of the variance-covariance matrices \(V^A=Cov\left(\hat{\beta}^A\right)\) and \(V^B =Cov\left(\hat{\beta}^B\right)\). In practice, with large enough samples, \(V^A\) and \(V^B\) can be estimated from the logistic regression and then \(S_q=\hat{V^q}=\widehat{Cov}(\hat{q})=\hat{V}^A+\hat{V}^B\).

The equivalence test should reject the null when the observed difference vector \(\hat{q}\) is small. We measure the size of the vector using the squared Mahalanobis norm \(\|v\|^2_{\Sigma} = v^T\Sigma^{-1}v\), where we set \(\Sigma = V^q\). The size of \(\hat{q}\) is therefore the Wald statistic, \(W =\hat{q}^T S_q^{-1} \hat{q}\); \(W\) follows a non-centralized chi-square distribution with non-centrality parameter \(\epsilon_\beta^2 = \|q\|^2_{V_q}\).
To set an equivalence threshold, we set an allowed difference per coefficient \(\delta_{\beta_i}\), then write \(\underaccent{\bar}{\delta}_\beta = (\delta_{\beta_0},...,\delta_{\beta_{p-1}})^T\in \mathbb{R}^p\).
We substitute \(V_q\) with its consistent estimator \(S_q\), the equivalence threshold can then be set to the Mahalanobis \(S_q\) size of \(\underaccent{\bar}{\delta}_{\beta}\)
\[\epsilon_\beta^2 = \|\underaccent{\bar}{\delta}_\beta\|^2_{S_q},\]
and the equivalence region is the set of vectors \(\left\{ k\in\mathbb{R}^p: \|k\|_{S_q} < \|\underaccent{\bar}{\delta}_\beta\|^2_{S_q}\right\}\).

Given the required equivalence threshold \(\epsilon_\beta^2\), we can identify the following hypotheses:
\[ H_0: \|\beta_A - \beta_B\|^2_{V^q} \geq \epsilon_\beta^2,\qquad H_1: \|\beta_A - \beta_B\|^2_{V^q} < \epsilon_\beta^2. \]
To ensure an \(\alpha\)-level test, we set the critical value to the \(\alpha\)-percentile of the non-centralized chi-squared distribution with non-centrality parameter \(\epsilon_\beta^2\):
\[ \beta_A \text{ is equivalent to } \beta_B \text{ if } W<\chi^2_{\alpha,p}(\epsilon_\beta^2). \]

\hypertarget{choosing-delta_beta}{%
\subsubsection{\texorpdfstring{Choosing \(\delta_\beta\)}{Choosing \textbackslash delta\_\textbackslash beta}}\label{choosing-delta_beta}}

As covariates might vary in scale and scientific importance, it is up to the researcher to specify an appropriate sensitivity level for each covariate. A possible alternative is scaling of each covariate, then either specifying the sensitivity level in terms of standard deviations or using a single sensitivity level for all covariates: \(\underaccent{\bar}{\delta}_\beta = (\delta_{\beta},...,\delta_{\beta})^T\).

\hypertarget{comparing-theta-hat}{%
\subsection{Individual Predictive Equivalence: Testing Log-Odds Vectors}\label{comparing-theta-hat}}

The response probabilities for each individual can be coded as odds. The logistic regression model has the odds for an individual \(i\) defined as \(\frac{P(y_i=1|x_i)}{P(y_i=0|x_i)}=e^{\theta_i}\). The odds ratio for \(x_1,x_2\) is then defined as \(\frac{e^{\theta_2}}{e^{\theta_1}}=e^{(x_2-x_1)^T\beta}\). This could also be used when comparing two estimates for the parameter vector (\(\hat{\beta}^A\) against \(\hat{\beta}^B\)): \(\frac{e^{\hat{\theta}_i^A}}{e^{\hat{\theta}_i^B}}=e^{x_i^T(\hat{\beta}^A-\hat{\beta}^B)}=e^{\xi_i}\).

Recall that individual predictive equivalence is achieved when two models yield similar predictions for a fixed set of observations. For logistic regression models, we compare the log-odds produced by the two models. The models obtain equivalence if the expectation of the absolute log-odd difference is significantly smaller than the predefined threshold.

Given a testing population \(X^{test}\) of size \(m\), we compare the observed log-odds obtained from models \(M^A\) and \(M^B\). Usually, \(X^{test}\) would be associated with one of the populations (without loss of generality, population A). In that case, we would consider the log-odds vector of \(\hat{\theta}^A_A\) to be the ``gold-standard'' for \(X^{test}\); the test assesses the effect of replacing \(\hat{\theta}^A_A\) with predictions obtained by applying model \(M^B\), which is based on a different population, to \(X^{test}\).
Note that the equivalence test does not privilege \(M_A\) or \(M_B\). However, our proposal for setting the equivalence threshold, we prefer using the gold-standard predictions (those of \(M_A\)).

In the following we assume that the observations in \(X^{test}\) are a simple random sample from our population of interest, and that they are independent from the observations used for training \(M^A\) and \(M^B\). The analysis is conditional on the two models (i.e.~on \(\hat{\beta}^A,\hat{\beta}^B\)), so that all variability is due to the sampling of \(X^{test}\).

For a sample \(x_i\in X^{test}, i=1,...,n\), model \(M^A\) yields log-odds \(\hat{\theta}_i^A\) and model \(M^B\) yields \(\hat{\theta}_i^B\). The odds ratio for \(x_i\) would be \(e^{\hat{\theta}^A_i-\hat{\theta}^B_i}=e^{\xi_i}\) and the absolute log-odds ratio is
\begin{equation}
{\xi}_i=\left|\hat{\theta}^A_i-\hat{\theta}^B_i\right|=\left|x^T_i\hat{q}\right|=\max\left\{\log\left(\frac{e^{\hat{\theta}_i^A}}{e^{\hat{\theta}_i^B}}\right),\log\left(\frac{e^{\hat{\theta}_i^B}}{e^{\hat{\theta}_i^A}}\right)\right\}.
\label{eq:log-odds-ratio}
\end{equation}
Conditional on \(\hat{\beta}^A,\hat{\beta}^B\), the values of the absolute log-odds ratio \(\xi_i\) are independent and identically distributed due to sampling of individuals to test set.
We denote this conditional distribution \(G\),
and let \(\mu_\xi\) and \(\sigma_\xi\) be its mean and standard deviation.
Denoting \(\epsilon_{\theta}\) as the allowed expected difference in log-odds ratio, we would like to verify whether \(\mu_\xi<\epsilon_{\theta}\) using the equivalence testing framework. Assuming the sample is large enough, we use a one-sided t-test on the absolute difference of log odds.

As we are interested in equivalence testing, the relevant hypotheses are
\[H_0:\mu_\xi\ge\epsilon_\theta,~~~H_1:\mu_\xi<\epsilon_\theta.\]

Although the conditional distribution \(G\) is unknown, we rely on the Central Limit Theorem to claim that the mean absolute log-odds value \(\bar{\xi}\) is normally distributed. As its variance is unknown, we should use a \(t\)-test. The level-\(\alpha\) equivalence test is:
\begin{equation}
\left\{ \frac{\sqrt{m}\left(\bar{\xi}-\epsilon_\theta\right)}{\sqrt{\widehat{Var}({\xi})}} < t_{\alpha,m-1} \right\}.
\label{eq:log-odds-test}
\end{equation}

\begin{figure}[H]

{\centering \includegraphics{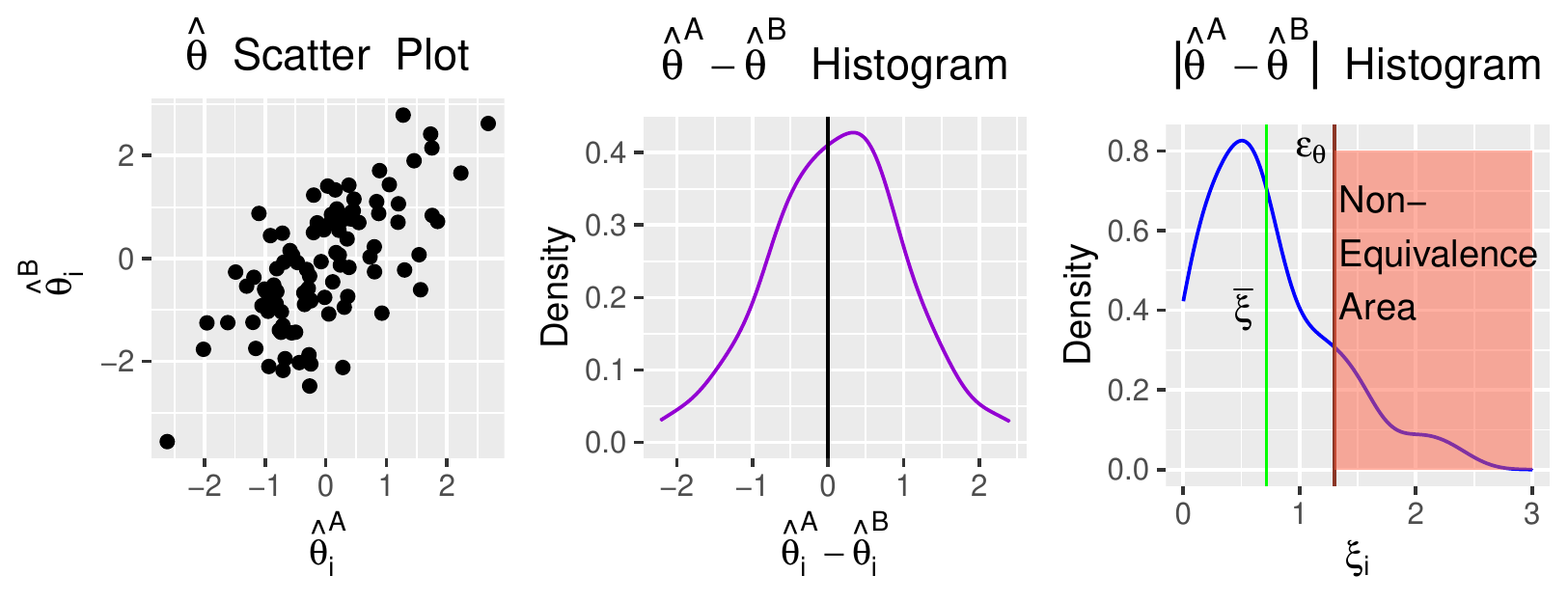} 

}

\caption{(Left) A scatter plot comparing the linear predictors from model B ($\hat{\theta}^B_i$) and model A ($\hat{\theta}^A_i$). (Center) the density of the corresponding difference in log odds ($\hat{\theta}^A_i-\hat{\theta}^B_i$). (Right) A density plot of the absolute differences $\xi_i=\left|\hat{\theta}^B_i-\hat{\theta}^B_i\right|$ including the equivalence threshold and the non-equivalence area. If $\bar{\xi} < \epsilon_{\theta}$, we would like the test to reject the inequivalence hypothesis $H_0$.}\label{fig:3part}
\end{figure}

Figure \ref{fig:3part} describes our procedure: Two sets of paired \(\hat{\theta}\) values provide us with a vector of log odd ratios and of absolute log odd ratios. The mean absolute log odds ratio is eventually compared against the equivalence threshold. Data was simulated using \(\hat{\theta}^A_i \sim \mathcal{N}(0,1)\) and \(\hat{\theta}^B_i \sim \mathcal{N}(\hat{\theta}^A_i,1)\).

\hypertarget{choosing-delta_theta}{%
\subsubsection{\texorpdfstring{Choosing \(\delta_\theta\)}{Choosing \textbackslash delta\_\textbackslash theta}}\label{choosing-delta_theta}}

To set the sensitivity level for the individual predictive equivalence \(\delta_\theta\), we propose looking at the distances between the estimated log odds and the classification threshold.

Consider vectors \(\hat{\theta}_A^A,\hat{\theta}_B^B\), as presented in Figure \ref{fig:pops1}. These are the predicted log-odds for test sets \(X^{test}_A,X^{test}_B\). These log-odds are obtained using models trained on data from the same population: Model \(M^A\) was trained using \(X^{train}_A\), and its predicted log-odds for \(X^{train}_A\) are \(\hat{\theta}_A^A\).

Consider \(\hat{\theta}\) to be such vector of \(m\) estimated log-odd values, without loss of generality we choose \(\hat{\theta}=\hat{\theta}_A^A\).
In a calibrated logistic regression, the classification of the \(i^{th}\) subject by the model (\(\hat{y}_i\)) is 1 when \(\hat{\theta}_i >0\) and 0 otherwise.
The absolute log-odds \(\left| \hat{\theta}_i \right|\) is minimal change to the log-odds that could change to the classification.
We therefore propose setting the equivalence threshold \(\epsilon_\theta\) to be a small quantile \(\delta_\theta\) (say \(\delta_\theta = 0.1\)) of the observed distribution of absolute log-odds
\[\epsilon_\theta =  \left| \hat{\theta}\right|_{(\lceil \delta_\theta \cdot m \rceil)}. \]

\begin{remarkun}
\normalfont One way to view the formulation of $\epsilon_\theta$ is that we set an upper bound on the fraction of subjects that would change in their classification if log-odd estimates of $M^A$ were replaced by the log-odd estimates of $M^B$. Note that this interpretation is not guaranteed to hold, because the equivalence test compares the mean absolute log-odds difference to $\epsilon_\theta$. If variation across individuals is large, we can still hypothetically get a larger fraction of individuals that change their classification.
\end{remarkun}

\begin{remarkun}
\normalfont We note that the definition of individual predictive equivalence does not depend on the linear logistic regression. Any two prediction models that output results in terms of probabilities or log-odds can be compared using this framework.
\end{remarkun}

\hypertarget{comparing-brier-score}{%
\subsection{Performance Equivalence : Testing Brier Scores}\label{comparing-brier-score}}

Sometimes, a sufficient criterion to retain a model fitted to different population is its overall performance in prediction. Two models obtain performance equivalence if the differences in their expected prediction loss is sufficiently small. To evaluate performance equivalence, we generate prediction values for the same test set from each model. These values are compared to expert-labeled or ground-truth outcomes using the Brier score. Similar to the previous section, the performance equivalence test is conditional on the fitted models.

The Brier score (Brier (\protect\hyperlink{ref-brier1950verification}{1950}), Benedetti (\protect\hyperlink{ref-benedetti2010scoring}{2010})) is defined as
\[BS=\frac{1}{m}\sum_{i=1}^{m}{(y_i-\hat{\pi}_i)^2}.\]

For a sufficiently large sample size, the CLT states that the sampling distribution of the Brier score converges to a normal distribution, with \(Var(BS)=\frac{1}{m}Var((y-\pi)^2)\) (\protect\hyperlink{ref-bradley2008sampling}{Bradley et al. 2008}). Consider two Brier scores \(BS^A,BS^B>0\) for models \(M^A,M^B\), calculated over the same dataset \(X^{test}\) of size \(m\). We would like to set equivalence thresholds for their ratio. Without loss of generality we choose the bound the ratio \(\frac{BS^B}{BS^A}\) with lower and upper thresholds \(0<\epsilon_L<1<\epsilon_U\), so that \(\epsilon_L<\frac{BS^B}{BS^A}<\epsilon_U\). The corresponding equivalence hypotheses are (\protect\hyperlink{ref-hauschke2007bioequivalence}{Hauschke et al. 2007}):
\[H_0:\frac{BS^B}{BS^A}\le\epsilon_L~~\text{or}~~\frac{BS^B}{BS^A}\ge\epsilon_U,\qquad H_1:\epsilon_L<\frac{BS^B}{BS^A}<\epsilon_U.\]
According to Hauschke et al. (\protect\hyperlink{ref-hauschke1999sample}{1999}), we reject the null hypothesis if
\[\frac{BS^B-\epsilon_L\cdot BS^A}{\sqrt{s_L^2}}>t_{\alpha,m-1} ~~~~\text{and}~~~~ \frac{BS^B-\epsilon_U\cdot BS^A}{\sqrt{s_U^2}}<-t_{\alpha,m-1}~~\]

where \(s_L^2\) is the sample variance of \(BS^B-\epsilon_L\cdot BS^A\) and \(s_U^2\) is the sample variance of \(BS^B-\epsilon_U\cdot BS^A\). We use these sample variances (rather than pooled variance) in order to compensate for possible correlation caused by both Brier scores computed for the same test set.

In addition to \(0<\epsilon_L<1<\epsilon_U\), we further assume \(\epsilon_U=\epsilon_L^{-1}=\epsilon_B\), similar to the 80/125 rule (see Chow and Liu (\protect\hyperlink{ref-chow2008design}{2008}) for introduction) widely used in bioequivalence. Our equivalence hypotheses can be written as
\[H_0:\frac{E[BS^B]}{E[BS^A]}\le\epsilon_B^{-1}~~\text{or}~~\frac{E[BS^B]}{E[BS^A]}\ge\epsilon_B,\qquad H_1:\epsilon_B^{-1}<\frac{E[BS^B]}{E[BS^A]}<\epsilon_B.\]
and the level-\(\alpha\) equivalence test is

\begin{equation}
\left\{ \frac{BS^B-\epsilon_B^{-1}\cdot BS^A}{\sqrt{\widehat{Var}(BS^B-\epsilon_B^{-1}\cdot BS^A)}}>t_{1-\alpha,m-1} ~~~~\text{and}~~~~ \frac{BS^B-\epsilon_B\cdot BS^A}{\sqrt{\widehat{Var}(BS^B-\epsilon_B\cdot BS^A)}}<-t_{1-\alpha,m-1} \right\}.
\label{eq:t-bsr-test}
\end{equation}

\hypertarget{choosing-delta_b}{%
\subsubsection{\texorpdfstring{Choosing \(\delta_B\)}{Choosing \textbackslash delta\_B}}\label{choosing-delta_b}}

To set the sensitivity level \(\delta_B\) for the performance equivalence, we suggest looking at the distances between the estimated probabilities and the real value of the dependent variable.

The ratio \(\frac{BS^B}{BS^A}\) reflects the change in Brier score caused by using model \(M^B\) on testing data from population \(A\). It may be more natural for the user to set the sensitivity level in terms of the absolute difference rather than the square difference. Hence, denoting \(\delta_B>1\) the acceptable score degradation or improvement, \(\left|y_i-\hat{\pi}^B_i\right|\le\delta_B\left|y_i-\hat{\pi}^A_i\right|\) or \(\left|y_i-\hat{\pi}^B_i\right|\le\frac{1}{\delta_B}\left|y_i-\hat{\pi}^A_i\right|\). That is, we take \(BS^A,BS^B\) as equivalent if \(\frac{BS^B}{BS^A}\in\left(\frac{1}{\delta_B^2},\delta_B^2\right)\). The relevant equivalence threshold is then \(\epsilon_B=\delta_B^2\).

\hypertarget{sim-stud}{%
\section{Simulation Study}\label{sim-stud}}

Our simulation demonstrates the performance of the three equivalence tests (descriptive equivalence, individual predictive equivalence and performance equivalence) for logistic regression models. We simulate data with different effects, sample sizes and effect sizes, then run the tests with different equivalence thresholds.

Our simulation is based on two populations \(A\) and \(B\), each characterised by a pair \((X,y)\) of a covariate matrix and a binary response vector. Descriptive equivalence is compared directly between the regression models, whereas individual predictive equivalence and performance equivalence are tested with respect to a third dataset \((X^{test},y^{test})\). Each equivalence test is compared to a standard null-hypothesis significance test. For each method, we measure the proportion for trials in which it had identified models \(M^A,M^B\) as equivalent (i.e rejecting the null in equivalence tests, not rejecting the null in significance tests).

The descriptive equivalence method is compared to the deviance test: Use the design matrices
\(X_{G_r}=\begin{pmatrix}1&X^A \\1& X^B\end{pmatrix}, X_G=\begin{pmatrix}1&1 & X^A & 0 \\1&0 & 0 & X^B\end{pmatrix}\) for the \emph{reduced} and the \emph{full} models respectively, with a gender indicator in the full model. For each of those, a logistic regression model is fitted and \(\hat{\beta}_{reduced},\hat{\beta}_{full}\) are found. The respective likelihood function values are denoted \(l(\hat{\beta}_{reduced}),l(\hat{\beta}_{full})\) and the test statistic is \(D=2(l(\hat{\beta}_{reduced})-l(\hat{\beta}_{full}))\). Denoting \(d\) as the difference in degrees of freedom between the models (in this case \(d=p\)), we get that \(D\sim\chi^2_d\).

The individual predictive equivalence method is compared to the Hosmer-Lemeshow test (\protect\hyperlink{ref-hosmer2013applied}{Hosmer Jr et al. 2013}): Given the samples \((x_i,y_i)\), we can classify them to \(G\) distinct groups according to the fitted probabilities \(\hat{\pi}_i\). Next, for each group \(g\in\{1,...,G\}\) we can calculate the following: \(n_g\) group size; \(O_{1g}\) the number of observed \(y_i=1\) events; \(E_{1g}\) the expected number of \(y_i=1\) events;
\(\bar{\pi}_g\) the average fitted probability. The Hosmer-Lemeshow test statistic is then calculated as
\(H=\sum_{g=1}^{G}{\left\{ (O_{1g}-E_{1g})^2 / (n_g\bar{\pi}_g(1-\bar{\pi}_g)\right\}}\). Using this test with \(G\) groups, we get that \(H\sim\chi^2_{G-2}\). The Hosmer-Lemeshow test is conducted using \texttt{ResourceSelection} (\protect\hyperlink{ref-R-ResourceSelection}{Lele et al. 2019}).

The performance equivalence method is compared to a \(t\)-test on the difference of the Brier scores. Using the terms defined in Section \ref{comparing-brier-score}, let \(BS_{A},BS_{B}\) be the Brier scores, we get that the two-sample \(t\)-test statistic is \(T=\frac{\sqrt{m}(BS_{B}-BS_{A})}{\sqrt{Var(BS_{B}-BS_{A})}}\)
and the level-\(\alpha\) \(t\)-test is \(\left\{|T|<t_{1-\alpha/2,m-1}\right\}\).

\textbf{Simulation Settings}

We sample \(X^A,X^B,X^{test}\) of length \(n\), where \(x^A_i, x^B_i,x^{test}_i \sim\mathcal{N}(0,I_p)\) and \(p=3\). For population \(A\) we set the regression coefficients to \(1\), so that the linear predictors are \(\theta^A_i=1+\sum_{j=1}^px^A_{ij}\), the probabilities \(\pi^A_i=\frac{e^{\theta_i^A}}{1+e^{\theta_i^A}}\) and the dependent variable \(y^A_i\sim Ber(\pi^A_i)\). We set the data for the \(test\) population in the same manner. We simulate different effects by setting \(\theta^B_i\) and \(\pi^B_i\) in different ways.

All simulations use \(n=\{100,200,...,1000, 1500, 2000, 2500, 5000, 7500, 10000\}\) and \(\alpha=0.05\), as well as the following sensitivity levels: \(\delta_\beta=\{0.1, 0.25, 0.5, 1\}\) for descriptive equivalence, \(\delta_\theta=\{0.025, 0.05, 0.1, 0.2\}\) for individual predictive equivalence and \(\delta_B=\{1.01, 1.05, 1.1, 1.2\}\) for performance equivalence. Each combination of sample size \(n\), effect type and effect size \(k\) is simulated 1000 times.

\hypertarget{log-odds-multiplicative-effect}{%
\subsection{Log-Odds Multiplicative Effect}\label{log-odds-multiplicative-effect}}

\begin{figure}[!h]

{\centering \includegraphics{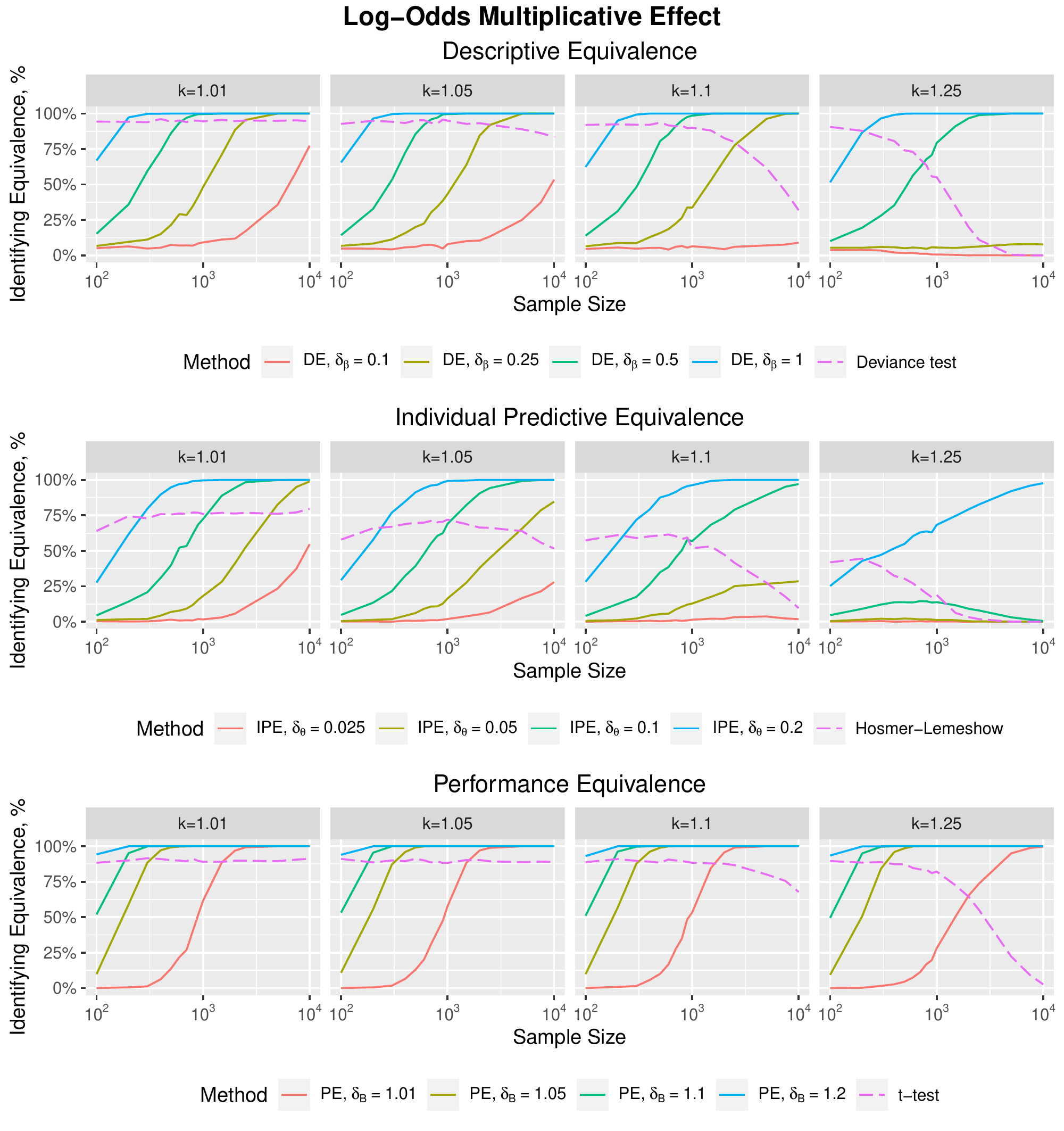} 

}

\caption{Testing the performance of different comparison methods against sample size, under log-odds multiplicative effect. DE stands for descriptive equivalence test (with different $\delta_\beta$ values), IPE stands for individual predictive equivalence test (with different $\delta_\theta$ values), PE stands for performance equivalence test (with different $\delta_B$ values).}\label{fig:sim1-multi}
\end{figure}

In this simulation we set a multiplicative effect on the log-odds, \(\theta^B_i=k_i\cdot\left(1+\sum_{j=1}^px^B_{ij}\right)\) with \(k_i\sim\mathcal{N}(k,0.1)\). We use effect sizes \(k=\{1.01, 1.05, 1.1, 1.25\}\). This effect is intended for assessing the sensitivity of the descriptive equivalence method. The results of this simulation are depicted in Figure \ref{fig:sim1-multi}.

As \(\theta^A_i=1+\sum_{j=1}^px^A_{ij}\), we expect the coefficients vector to have the form \(q=(k-1,...,k-1)\). We can see that the descriptive equivalence (DE) method identifies equivalence in most cases when using \(\delta_\beta>k-1\) (that is, sensitivity level larger than the actual effect) and as the sample size is large enough. On the other hand, when using \(\delta_\beta<k-1\) it fails to identify equivalence. The deviance test fails to identify equivalence (i.e not rejecting the null hypothesis) for large effect size and large sample size.

As the three equivalence methods have a cascading form, we expect that identification of descriptive equivalence (DE) would also imply the identification of individual predictive equivalence (IPE) and performance equivalence (PE), upon choosing appropriate equivalence thresholds. We can see that this is the case for both IPE and PE. We can also see that the equivalence methods' ratio of identifying equivalence (using appropriate thresholds) increases as sample size grows, while the ratio of identifying equivalence for the corresponding significance tests decreases. Overall, for large effects and/or sample sizes, the equivalence-based methods perform better than the significance-based methods (in terms of successfully identifying model equivalence).

\hypertarget{log-odds-additive-effect}{%
\subsection{Log-Odds Additive Effect}\label{log-odds-additive-effect}}

In this simulation we set a additive effect on the log-odds, \(\theta^B_i=k_i+\left(1+\sum_{j=1}^px^B_{ij}\right)\) with \(k_i\sim\mathcal{N}(k,0.1)\). We use effect sizes \(k=\{0.05, 0.1, 0.25, 0.5\}\). This effect is intended for assessing the sensitivity of the individual predictive equivalence method. The results of this simulation are depicted in Supplementary Figure \ref{fig:sim2-add}.

As \(\theta^A_i=1+\sum_{j=1}^px^A_{ij}\), we expect the log-odds ratio to have the form \(\xi_i=k\). We can see that the individual predictive equivalence (IPE) method identifies equivalence in most cases when using a large \(\delta_\theta\) values. On the other hand, when using small \(\delta_\theta\) values it fails to identify equivalence. The Hosmer-Lemeshow test fails to identify equivalence (i.e not rejecting the null hypothesis) for almost all large effect sizes and large sample sizes. Due to the cascading form of the equivalence methods, we expect that identification of individual predictive equivalence (IPE) would also imply the identification of performance equivalence (PE), upon choosing appropriate equivalence thresholds. We can see that this is indeed the case.

\begin{figure}[!h]

{\centering \includegraphics{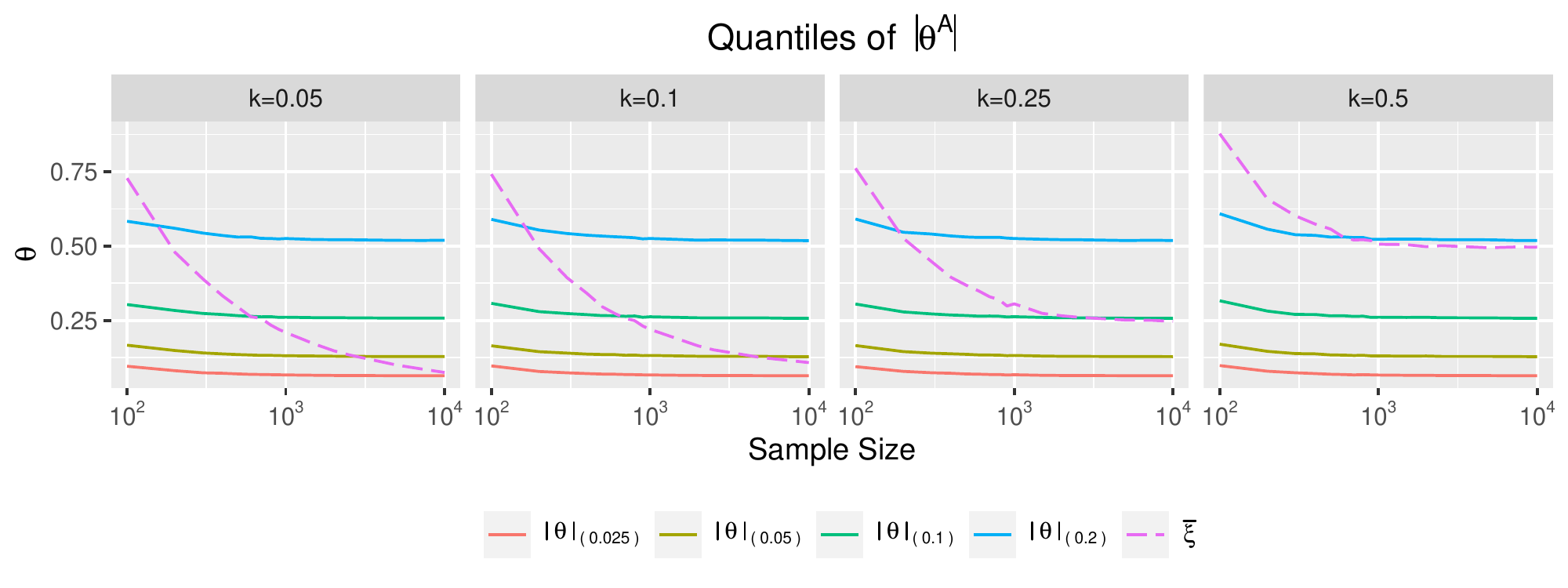} 

}

\caption{Different quantiles of $|\theta^A|$ and the MAD of $\theta^B-\theta^A$.}\label{fig:sim2-thetas}
\end{figure}

The additive effect allows us to compare the equivalence boundaries against the actual MAD of \(\xi=\theta^B-\theta^A\), as depicted in Figure \ref{fig:sim2-thetas}. As sample size grows, all the quantiles of \(|\theta^A|\) converge, as well as \(\bar{\xi}\). The limit of \(\bar{\xi}\) is the effect size \(k\), as expected. For a large effect size such as \(k=0.5\), we should choose a very lenient equivalence boundary for identifying individual predictive equivalence (in the above example equivalence is identified if we let 20\% of the samples flip their prediction).

\hypertarget{probability-multiplicative-effect}{%
\subsection{Probability Multiplicative Effect}\label{probability-multiplicative-effect}}

In this simulation we set a multiplicative effect on the predicted probabilities. Let \(\theta^B_i=1+\sum_{j=1}^px^B_{ij}\), we set \(\pi^B_i=k_i\cdot \frac{e^{\theta^B_i}}{1+e^{\theta^B_i}}\) with \(k_i\sim\mathcal{N}(k,0.1)\). We use effect sizes \(k=\{1.01, 1.1, 1.25, 1.5\}\).
For the (very rare) cases where \(\pi^B_i>1\) we set \(\pi^B_i=1-10^{-6}\) and for \(\pi^B_i<0\) we set \(\pi^B_i=10^{-6}\). This effect is intended for assessing the sensitivity of the performance equivalence method.
The results of this simulation are depicted in Supplementary Figure \ref{fig:sim3-prob}.

Using a large enough equivalence threshold, the performance equivalence method (PE) successfully identifies equivalence even for large effect sizes. This is not the case with the descriptive equivalence (DE) and individual predictive equivalence (IPE) methods, which fail to identify equivalence under large effect sizes. The \(t\)-test fails to identify equivalence (i.e not rejecting the null hypothesis) for large effect sizes and large sample sizes.

\hypertarget{errtc}{%
\subsection{Error Rate Control}\label{errtc}}

In this simulation we test the three equivalence methods for their error rate control, under two conditions: no effect at all (equal models) and a log-odds multiplicative effect of size 1.5 (unequal models). Each condition is simulated 1000 times for \(n=\{100, 1000, 10000\}\) and \(\alpha=\{0.05, 0.1\}\), using the following sensitivity levels: 0.2 for the descriptive equivalence method, 0.05 for the individual predictive equivalence method and 1.005 for the performance equivalence method. These levels were chosen for being fairly strict. The results appear in Table \ref{tab:errtc-table-no}. We can see that under no effect (equal models), the three equivalence methods identify model equivalence when the sample size is large enough. On the other hand, we see that for unequal models the error rate of the methods is kept below \(\alpha\).

\begin{table}

\caption{\label{tab:errtc-table-no}Error Rates}
\centering
\begin{threeparttable}
\begin{tabular}[t]{lrrr|>{}rrr|>{}rrr|>{}rrr}
\toprule
\multicolumn{1}{c}{ } & \multicolumn{6}{c}{Equal Models} & \multicolumn{6}{c}{Unequal Models} \\
\cmidrule(l{3pt}r{3pt}){2-7} \cmidrule(l{3pt}r{3pt}){8-13}
\multicolumn{1}{c}{ } & \multicolumn{3}{c}{$\alpha=0.05$} & \multicolumn{3}{c}{$\alpha=0.1$} & \multicolumn{3}{c}{$\alpha=0.05$} & \multicolumn{3}{c}{$\alpha=0.1$} \\
\cmidrule(l{3pt}r{3pt}){2-4} \cmidrule(l{3pt}r{3pt}){5-7} \cmidrule(l{3pt}r{3pt}){8-10} \cmidrule(l{3pt}r{3pt}){11-13}
  & DE & IPE & PE & DE & IPE & PE & DE & IPE & PE & DE & IPE & PE\\
\midrule
$n=100$ & 0.06 & 0.01 & 0.00 & 0.10 & 0.01 & 0.00 & 0.02 & 0 & 0 & 0.04 & 0.01 & 0\\
$n=1000$ & 0.26 & 0.16 & 0.12 & 0.46 & 0.18 & 0.13 & 0.00 & 0 & 0 & 0.00 & 0.00 & 0\\
$n=10000$ & 1.00 & 0.98 & 1.00 & 1.00 & 0.99 & 1.00 & 0.00 & 0 & 0 & 0.00 & 0.00 & 0\\
\bottomrule
\end{tabular}
\begin{tablenotes}
\item[*] DE stands for Descriptive Equivalence, IPE for Individual Predictive Equivalence, 
 PE for Performance Equivalence.
\end{tablenotes}
\end{threeparttable}
\end{table}

\hypertarget{matal}{%
\section{Learning Disabilities and MATAL Data Results}\label{matal}}

A learning disability is defined as a ``neurodevelopmental disorder {[}\ldots{]} characterized by persistent and impairing difficulties with learning foundational academic skills in reading, writing, and/or math'' (\protect\hyperlink{ref-american2013diagnostic}{American Psychiatric Association and others 2013}). MATAL (Ben-Simon (\protect\hyperlink{ref-ben2007matal}{2007}), Ben-Simon et al. (\protect\hyperlink{ref-ben2008regulating}{2008}), Ben-Simon (\protect\hyperlink{ref-ben2013matal}{2013})) is a computer-based test battery for the diagnosis of learning disabilities (dyslexia, dysgraphia, and dyscalculia) and Attention Deficit \& Hyperactivity Disorder (ADHD) for applicants to Israeli higher education institutions and for currently enrolled students. MATAL was developed by the Israeli National Institute for Testing \& Evaluation in cooperation with the Israeli Council of Higher Education, as part of an endeavor to develop policy and procedure for standardizing and regulating the diagnosis of learning disabilities in higher education and the provision of test accommodations. Since its inauguration in 2007, the MATAL system has diagnosed more than 40,000 enrolled students and applicants to higher education (around 5,000 yearly). It became the standard diagnosis tool for most Israeli higher education institutions. The MATAL system consists of 22 standardized neuropsychological tests, whose results are summarized into 10 functioning scores. The system provides its predictions regarding each disability using logistic regression models, with the functioning scores as covariates. Experts are trained to incorporate the computerized diagnosis with the various test scores and the background information provided in reaching their decision.

A system such as MATAL provides a prime example where any change to the system would involve a great scientific cost: Experts have been trained and gained experience incorporating results of the system in their decision making process. The diagnosis data collected has been used for further neuropsychological research regarding learning disabilities and cognitive functioning. Changing the model might impair the reliability of previous diagnoses and long-term studies. It will also require the experts to relearn how to interpret the outcomes. Therefore, any change to the underlying diagnosis model would involve both scientific and inter-organization costs. We will examine the decision-making process when determining whether a new population is similar enough for there to be no need of re-estimating the model.

We will use the MATAL data to study the regression models for two sub-populations: male and female students. For two disabilities, dysgraphia and dyscalculia, we will fit separate models for each sub-population. We will then use the different Equivalence tests to verify or refute the equivalence of the two populations for these disabilities.

\hypertarget{data-generation}{%
\subsection*{Data Generation ~}\label{data-generation}}
\addcontentsline{toc}{subsection}{Data Generation ~}

The original MATAL data incorporates personal and proprietorial information, so the following results are based on data regenerated according to the estimated distributions. For each response and each gender, we estimated the joint distribution of 5 functioning scores using a Gaussian copula (\protect\hyperlink{ref-joe2014dependence}{Joe 2014}) with Gamma marginal distributions. The full regeneration process is explained in Section \ref{matal-dg}. We resampled data points for each category, using \(n_{female}^{train}=n_{male}^{train}=3000\) for two disabilities: dysgraphia and dyscalculia.

Each sub-sample was used to construct logistic regression models for dysgraphia and dyscalculia. The prediction of each disability is made using 10 functioning scores as predictors. Each disability uses different predictors, according to their relevance (e.g.~quantitative skill is used only for predicting dyscalculia, whereas verbal fluency is used only for predicting dysgraphia). The following comparisons relate to multiple combinations of disability and gender. We use \(\alpha=0.05\) and sensitivity levels \(\delta_\beta=0.1\) for the descriptive equivalence, \(\delta_\theta=7.5\%\) for the individual predictive equivalence method and \(\delta_B=1.1\) for the performance equivalence. The individual predictive equivalence and performance equivalence methods were tested with resampled testing datasets (\(n_{female}^{test}=n_{male}^{test}=1000\)).

\hypertarget{results}{%
\subsection*{Results ~}\label{results}}
\addcontentsline{toc}{subsection}{Results ~}

The coefficients for the female and male dysgraphia models are presented in Table \ref{tab:dysg-stats}, whereas the coefficients for dyscalculia are in Table \ref{tab:dysc-stats}. The Brier scores for all models are presented in Table \ref{tab:brier-table}. Dysgraphia models were found to be equivalent: The descriptive equivalence test, the individual predictive equivalence test and the performance equivalence test have found the models equivalent for the female data, as well as the \(t\)-test for the Brier scores. For Dyscalculia, we found only performance equivalence. A brief discussion of the results follows, along with their compliance to previous findings. The full result are found in Tables \ref{tab:dysg-table-print} and \ref{tab:dysc-table-print}.

\begin{table}

\caption{\label{tab:dysg-stats}Coefficients for Dysgraphia Models}
\centering
\begin{tabular}[t]{lrrrr}
\toprule
  & (Intercept) & $x_1$ & $x_2$ & $x_3$\\
\midrule
Female model & -3.022 & -0.090 & 0.248 & -0.263\\
Male model & -3.261 & -0.049 & -0.042 & -0.197\\
\midrule
$\hat{q}$ & 0.240 & -0.042 & 0.290 & -0.066\\
\bottomrule
\end{tabular}
\end{table}

\begin{table}

\caption{\label{tab:dysc-stats}Coefficients for Dyscalculia Models}
\centering
\begin{tabular}[t]{lrrr}
\toprule
  & (Intercept) & $x_4$ & $x_5$\\
\midrule
Female model & -3.189 & -0.430 & -7.027\\
Male model & -2.121 & -0.387 & -5.614\\
\midrule
$\hat{q}$ & -1.068 & -0.043 & -1.414\\
\bottomrule
\end{tabular}
\end{table}

\begin{table}

\caption{\label{tab:brier-table}Brier Scores for Dysgraphia and Dyscalculia Models}
\centering
\begin{tabular}[t]{lrr|>{}rr}
\toprule
\multicolumn{1}{c}{ } & \multicolumn{2}{c}{Dysgraphia} & \multicolumn{2}{c}{Dyscalculia} \\
\cmidrule(l{3pt}r{3pt}){2-3} \cmidrule(l{3pt}r{3pt}){4-5}
  & Male Model & Female Model & Male Model & Female Model\\
\midrule
Male Data & 0.1060 & 0.1164 & 0.1103 & 0.1259\\
Female Data & 0.1216 & 0.1086 & 0.1133 & 0.1066\\
\bottomrule
\end{tabular}
\end{table}

\begin{table}[!h]

\caption{\label{tab:dysg-table-print}Summary of Tests for Dysgraphia Models}
\centering
\begin{threeparttable}
\begin{tabular}[t]{lllccccc}
\toprule
Method & Type & Test Set & $\epsilon$ & $C_\alpha$ & Test Stat. & P-value & Models Differ?\\
\midrule
DE ($\delta_\beta=0.1$) & Equiv. & - & - & 437.148 & 87.481 & 0.000 & No\\
Deviance Test & Signif. & - & - & 9.488 & 91.548 & 0.000 & Yes\\
\midrule
IPE ($\delta_\theta=7.5\%$) & Equiv. & Female & 0.528 & -1.646 & -1.983 & 0.024 & No\\
Hosmer-Lemeshow & Signif. & Female & - & 15.507 & 255.914 & 0.000 & Yes\\
\midrule
IPE ($\delta_\theta=7.5\%$) & Equiv. & Male & 0.435 & -1.646 & 14.464 & 1.000 & Yes\\
Hosmer-Lemeshow & Signif. & Male & - & 15.507 & 552.907 & 0.000 & Yes\\
\midrule
PE ($\delta_B=1.1$), $t_L$ & Equiv. & Female & 1.21 & 1.962 & 2.869 & 0.002 & No\\
PE ($\delta_B=1.1$), $t_U$ & Equiv. &  &  & -1.962 & -10.150 & 0.000 & \\
Brier $t$-test & Signif. & Female & - & 1.646 & -7.718 & 0.000 & Yes\\
\midrule
PE ($\delta_B=1.1$), $t_L$ & Equiv. & Male & 1.21 & 1.962 & 8.303 & 0.000 & No\\
PE ($\delta_B=1.1$), $t_U$ & Equiv. &  &  & -1.962 & -3.195 & 0.001 & \\
Brier $t$-test & Signif. & Male & - & 1.646 & 11.410 & 0.000 & Yes\\
\bottomrule
\end{tabular}
\begin{tablenotes}
\item[*] DE stands for Descriptive Equivalence, IPE for Individual Predictive Equivalence, 
 PE for Performance Equivalence.
\end{tablenotes}
\end{threeparttable}
\end{table}

\begin{table}[!h]

\caption{\label{tab:dysc-table-print}Summary of Tests for Dyscalculia Models}
\centering
\begin{threeparttable}
\begin{tabular}[t]{lllccccc}
\toprule
Method & Type & Test Set & $\epsilon$ & $C_\alpha$ & Test Stat. & P-value & Models Differ?\\
\midrule
DE ($\delta_\beta=0.1$) & Equiv. & - & - & 11.783 & 112.715 & 1.000 & Yes\\
Deviance Test & Signif. & - & - & 7.815 & 120.510 & 0.000 & Yes\\
\midrule
IPE ($\delta_\theta=7.5\%$) & Equiv. & Female & 0.273 & -1.646 & 74.159 & 1.000 & Yes\\
Hosmer-Lemeshow & Signif. & Female & - & 15.507 & 345.777 & 0.000 & Yes\\
\midrule
IPE ($\delta_\theta=7.5\%$) & Equiv. & Male & 0.338 & -1.646 & 93.567 & 1.000 & Yes\\
Hosmer-Lemeshow & Signif. & Male & - & 15.507 & 604.883 & 0.000 & Yes\\
\midrule
PE ($\delta_B=1.1$), $t_L$ & Equiv. & Female & 1.21 & 1.962 & 4.366 & 0.000 & No\\
PE ($\delta_B=1.1$), $t_U$ & Equiv. &  &  & -1.962 & -9.463 & 0.000 & \\
Brier $t$-test & Signif. & Female & - & 1.646 & 3.406 & 0.000 & Yes\\
\midrule
PE ($\delta_B=1.1$), $t_L$ & Equiv. & Male & 1.21 & 1.962 & 9.518 & 0.000 & No\\
PE ($\delta_B=1.1$), $t_U$ & Equiv. &  &  & -1.962 & -2.759 & 0.003 & \\
Brier $t$-test & Signif. & Male & - & 1.646 & -0.419 & 0.338 & No\\
\bottomrule
\end{tabular}
\begin{tablenotes}
\item[*] DE stands for Descriptive Equivalence, IPE for Individual Predictive Equivalence, 
 PE for Performance Equivalence.
\end{tablenotes}
\end{threeparttable}
\end{table}

\textbf{Dysgraphia}

For dysgraphia, we find a repeating pattern in which the equivalence tests find equivalence, whereas the
usual significance tests sometimes rejected the null-hypothesis of no-difference.

The descriptive equivalence test for the selected \(\delta_\beta\) and \(\alpha\) values does reject the null hypothesis \(H_0: \left\|\hat{\beta}^{male} - \hat{\beta}^{female}\right\|_{\Sigma}\geq\lambda_\beta\), meaning the models describe dysgraphia in an equivalent manner. The Mahalanobis distance between the dysgraphia coefficient vectors is 87.5, while the allowed distance for our choice of \(\delta_{\beta}=0.1\) is 505.5.
The deviance test for the gender variable rejects the hypothesis ``\(x_{gender}\) does not improve the model'', meaning the models differ for our choice of \(\alpha=0.05\).

When testing the male model's fit for the female data, we observe differences between the significance test and equivalence test: The individual predictive equivalence test finds the models equivalent for the female data, while Hosmer-Lemeshow goodness-of-fit test rejects the null hypothesis (that the male model fits the female data). Our choice of \(\delta_{\theta}=7.5\)\% yields \(\epsilon_{\theta}= 0.528\) while \(\bar{{\xi}}= 0.498\). This means a negative numerator and a negative test statistic, so the inequivalence hypothesis is rejected.

When testing the female model's fit for the male data, both the individual predictive equivalence test and the Hosmer-Lemeshow goodness-of-fit test have found the models inequivalent. We got \(\delta_{\theta}= 0.435\) and \(\bar{{\xi}}= 0.678\). This means our choice of \(\delta_\theta=7.5\)\% is too strict for the male data.

The performance equivalence test has found the models equivalent for both male and female testing datasets, while the \(t\)-test for the Brier scores did not find find any equivalence.
We have set \(\delta_B=1.1\) as the sensitivity level for the performance equivalence test, meaning we tolerate Brier ratios in \([0.826,1.21]\). The observed Brier ratio \(\left(\frac{BS^{male}}{BS^{female}}\right)\) is 1.12 for the female test data and by 0.911 for the male test data. It comes as no surprise that the Brier scores were found equivalent in both cases.

Overall, it seems that the models are equivalent in describing the phenomenon of dysgraphia and perform equivalently in terms of prediction for the female data.

Berninger and O'Malley May (\protect\hyperlink{ref-berninger2011evidence}{2011}) has found males to be ``consistently more impaired'' than females in orthographic skills. The latter is known as a key factor in dysgraphia. fMRI studies in that research have found gender difference in brain activation only in regions associated with orthographic processing. However, no gender differences in brain activation were observed on other writing tasks. This might imply some gender similarities in dysgraphia, which explains the descriptive equivalence and predictive equivalence achieved for the female data.

\textbf{Dyscalculia}

The descriptive equivalence test for the selected \(\delta_\beta\) and \(\alpha\) values does reject the null hypothesis \(H_0: \left\|\hat{\beta}^{male} - \hat{\beta}^{female}\right\|_{\Sigma}\geq\lambda_\beta\), meaning the models describe dyscalculia in a different manner. The deviance test for the gender variable does not reject the hypothesis ``\(x_{gender}\) does not improve the model'', meaning the models differ for our choice of \(\alpha=0.05\).

The individual predictive equivalence test finds the models ineqiuvalent for both the female and male data. This means our choice of \(\delta_{\theta}=7.5\)\% might be too strict for the data.
When testing the male model's fit for the female data, the Hosmer-Lemeshow goodness-of-fit test rejects the null hypothesis (that the male model fits the female data). The same result is obtained when testing the female model's fit for the male data.

The performance equivalence tests for both the male and female datasets has found the models equivalent, as well as the \(t\)-test for the Brier scores over the male test data. This means our choice of \(\delta_B=1.1\)\% might be too lenient for the data.

Overall, it seems that the models are describing the phenomenon of dyscalculia in a different manner and differ in their individual predictions. The equivalence of Brier scores might be caused by the limited range of Brier scores.

The gender inequivalence for dyscalculia complies with Devine et al. (\protect\hyperlink{ref-devine2013gender}{2013}), which found that using discrepancy thresholds ``significantly more girls than boys could be defined as having developmental dyscalculia''.

\hypertarget{discussion}{%
\section{Discussion}\label{discussion}}

This work addresses the problem of assessing the suitability of using a single logistic regression model for different populations. Although no two populations are identical, fitting a different model for each population requires considerable scientific and administrative costs. Because identifying equivalent models is the goal of this investigation, we argue that the burden should be on the scientist to show that the models are equivalent up to a specified sensitivity level. We develop an equivalence testing framework for logistic regression models fitted to two different populations. This framework consists of three different equivalence tests: (a) descriptive equivalence, which compares the models' coefficient vectors; (b) individual predictive equivalence, which compares the models' log-odds estimate vectors; and (c) performance equivalence, which compares the models' average prediction accuracy using Brier scores. The proposed tests are based on asymptotic normality and are formed by adding equivalence regions and inverting the direction of the rejection regions.

The usage of the proposed methods is subject to the research goal of the scientist: if it is to show that the models describe a certain phenomenon in a similar manner, then the descriptive equivalence is the appropriate method; if it is to show that the models produce equal predictions for a given dataset, then the individual predictive equivalence is the appropriate method; if it is to show that the overall prediction accuracy is similar, then the performance equivalence is the appropriate method.

The three methods also form a cascade, as discussed in Section \ref{DSN}. Achieving a certain level of equivalence implies `weaker' forms of equivalence (see Section \ref{cascade}). On the other hand, finding two models inequivalent (that is, not rejecting the null hypothesis of an equivalence test) does not necessarily mean that `weaker' forms of equivalence are unobtainable. An example of such case can be found in Section \ref{matal}: The male and female models for diagnosing dyscalculia only achieve performance equivalence.

These three methods aren't a full framework, and more comparison steps should be suggested. Currently it is up for the researcher to decide what is an acceptable equivalence threshold, which was identified as a ``key methodological issue'' of equivalence testing by Greene et al. (\protect\hyperlink{ref-greene2008noninferiority}{2008}). When using more than one method, it might be advised to apply a multiplicity correction to the significance level, such as the Bonferroni correction or the Holm-Bonferroni method.

The methods described in this paper could be compared to some recent work: Unlike the equivalence methodology proposed by Dette et al. (\protect\hyperlink{ref-dette2018equivalence}{2018}), we use asymptotic parametric tests in this work (rather than bootstrap-based). In addition, the tests suggested use different metrics: The descriptive equivalence and individual predictive equivalence methods compare models by using Mahalanobis and \(L^1\)-distances (respectively) rather than \(L^2\) or \(L^\infty\). Moreover, we show cascading order between the distances we use.

The Mantel-Haenszel test counts cases according to the predicted binary output versus the actual output, much like the accuracy score \(\frac{1}{m}\sum_{i=1}^{m}{I\{\hat{y}_i=y_i\}}\). One advantage of the performance equivalence method over the Mantel-Haenszel equivalence (\protect\hyperlink{ref-casabianca2018statistical}{Casabianca and Lewis 2018}) is the usage of proper scoring rule, and another possible advantage is the use of a parametric test.

For individual predictive equivalence, we proposed a method for choosing the threshold by looking at the fitted data. Although this choice is a bit unorthodox, we believe it allow greater flexibility for the investigator. Moreover, as differences of logits or Brier scores have no meaning outside context (unlike boundaries set for the descriptive equivalence method), using percentile-based boundaries make these methods usable even for investigators who are not very familiar with statistical methodology. It is much easier to state ``I am willing to let 5\% of the samples flip their prediction'' than to use an allowed difference in logits, or use a pre-set boundary. The same applies for the performance equivalence method - ``I will tolerate deviations of up to 20\% in the Brier scores'' is easier that specifying a value.

One limitation of this work is that it is based on the normal approximation of the logistic regression estimates for the coefficients and the log-odds. Another one is the normal approximation of the Brier score, based on its MSE structure, although its full sampling distribution is known. Section \ref{matal} results also suggest that the log-odds equivalence method might be too strict. The proposed equivalence methods are compared to very common significance tests, while there might be other comparable methods out there with better performance.

Despite these drawbacks, this work does present a new approach towards to comparison of logistic regression models: The combination of different methods offers a comprehensive view of the models, and the usage of equivalence testing encourages using large sample sizes. Even as is, this framework could help researchers perform meta-analysis of results, as measurement invariance is used for factor analysis results (for example Byrne et al. (\protect\hyperlink{ref-byrne1989testing}{1989}); Collins and Gleaves (\protect\hyperlink{ref-collins1998race}{1998})). The ideas presented in this work could be used for some possible further developments:
(a) introduction of new comparison methods between logistic regression models, such as comparison of the outputted probabilities;
(b) incorporating equivalence tests in the context of assessing measurement invariance between two factor analysis models;
(c) extending the proposed methods for comparison of other classifier types;
(d) using the proposed methods in the psychometric context to provide new insights regarding differential item functioning (DIF).

\hypertarget{acknowledgements}{%
\section*{Acknowledgements ~}\label{acknowledgements}}
\addcontentsline{toc}{section}{Acknowledgements ~}

The authors would like to thank Anat Ben-Simon and Yoel Rapp for their long-term support of this research, as well as Henry Braun for helpful discussions. The authors would also like to express their gratitude towards three reviewers, who made constructive comments on an earlier version of this work.

Equivalence methods conducted using \texttt{LogRegEquiv} (\protect\hyperlink{ref-R-LogRegEquiv}{Ashiri-Prossner 2022}). Figures created using \texttt{ggplot2} (\protect\hyperlink{ref-R-ggplot2}{Wickham et al. 2019}), \texttt{gridExtra} (\protect\hyperlink{ref-R-gridExtra}{Auguie 2017}) and \texttt{latex2exp} (\protect\hyperlink{ref-R-latex2exp}{Meschiari 2021}). Tables were created using \texttt{kableExtra} (\protect\hyperlink{ref-R-kableExtra}{Zhu 2019}). Document authoring using \texttt{rmarkdown} (\protect\hyperlink{ref-R-rmarkdown}{Allaire et al. 2019}), \texttt{knitr} (\protect\hyperlink{ref-R-knitr}{Xie 2019a}) and \texttt{bookdown} (\protect\hyperlink{ref-R-bookdown}{Xie 2019b}).

\newpage

\hypertarget{supplements}{%
\section{Supplements}\label{supplements}}

\hypertarget{cascade}{%
\subsection{Cascading Order of Equivalence}\label{cascade}}

\begin{propun}
Descriptive equivalence implies individual predictive equivalence, using a proper equivalence threshold.
\end{propun}

\emph{Proof.}
Let models \(M^A,M^B\) achieve descriptive equivalence. That is, there exists some \(\epsilon_\beta>0\) for which \(\|\hat\beta^A - \hat\beta^B\|^2_{S_q} < \epsilon_\beta^2\), where \(\hat q=\hat\beta^A - \hat\beta^B\) and \(S_q=Cov\left(\hat q\right)\).
We can write \(\left|\hat{\theta}^A_i-\hat{\theta}^B_i\right|=\left|x^T_i\hat{q}\right|\). Given \(S_q\), we denote \(\lambda_1>...>\lambda_p\) as its eigenvalues. We can then bound the squared Mahalanobis norm using the \(L^2\) norm: Let \(\mathbf{c}\in\mathbb{R}^p\), we get \(\|\mathbf{c}\|^2\le\lambda_1\|\mathbf{c}\|_{S_q}^2\) (\protect\hyperlink{ref-jensen1997bounds}{Jensen 1997}). Using the power norm inequality we get
\[|\mathbf{c}|\le\|\mathbf{c}\|\le\sqrt{\lambda_1}\|\mathbf{c}\|_{S_q}\]
then using the Cauchy-Schwarz inequality, we get
\[\left|x^T_i\hat{q}\right|\le\|x_i\|\cdot\|\hat{q}\|\le \|x_i\|\sqrt{\lambda_1}\left\|\hat{q}\right\|_{S_q}\le\epsilon_\beta\sqrt{\lambda_1} \left\|x_i\right\|.\]

Using Jensen's inequality we can write
\[\left|\theta^A_i-\theta^B_i\right|=\left|E\left[x^T_i\hat{q}\right]\right|\le E\left[\left|x^T_i\hat{q}\right|\right]\le  \epsilon_\beta\sqrt{\lambda_1} E\left[\left\|x_i\right\|\right].\]

The function \(g(x)=x^2\) is strictly convex, so using Jensen's inequality we can write \(E\left[\left\|x_i\right\|\right]^2 \le E\left[\left\|x_i\right\|^2\right]\) and finally

\[E\left[\left|\theta^A_i-\theta^B_i\right|\right]\le\epsilon_\beta\sqrt{\lambda_1}\sqrt{E\left[\left\|x_i\right\|^2\right]}.\]

Assuming test set \(X\) of size \(m\), denote \(\mu_X\) as its expected value vector and \(\Sigma_X\) as its covariance matrix, we get \[E\left[\left|\theta^A-\theta^B\right|\right]\le \epsilon_\beta\sqrt{\lambda_1}\sqrt{{\mu_X}^T\mu_X+tr(\Sigma_X)},\]
meaning individual predictive equivalence is achieved. \(\blacksquare\)

\vspace{\baselineskip}

\begin{propun}
Individual predictive equivalence implies performance equivalence, using a proper equivalence threshold.
\end{propun}

\emph{Proof.}
Assume test set \(X_{test}\) of size \(m\) and let models \(M^A,M^B\) achieve individual predictive equivalence with respect to \(X_{test}\). That is, for a given significance level \(\alpha\) there exists some \(\epsilon_\theta>0\) for which the test in Equation (\ref{eq:log-odds-test}) does reject the null hypothesis.
This means we can bound the mean absolute difference of log-odds \(E\left[ |\xi_i|\right]<\epsilon_\theta\), which can also be written as \(E\left[\left|logit(\pi^B_i)-logit(\pi^A_i)\right|\right]<\epsilon_\theta\). The difference of logits can be written as
\[E\left[\left|(\log(\pi^B_i)+\log(1-\pi^A_i))-(\log(\pi^A_i)+\log(1-\pi^B_i))\right|\right]=E\left[\left|logit(\pi^B_i)-logit(\pi^A_i)\right|\right]<\epsilon_\theta\]
Using reverse triangle inequality
\[E\left[\left|\left|\log(\pi^B_i)+\log(1-\pi^A_i)\right|-\left|\log(\pi^A_i)+\log(1-\pi^B_i)\right|\right|\right]<E\left[\left|logit(\pi^B_i)-logit(\pi^A_i)\right|\right].\]
As both \(\log(\pi^B_i),\log(1-\pi^A_i)\) are negative, we get
\[E\left[\left|\left|\log(\pi^B_i)\right|-\left|\log(\pi^A_i)+\log(1-\pi^B_i)\right|\right|\right]<E\left[\left|\left|\log(\pi^B_i)+\log(1-\pi^A_i)\right|-\left|\log(\pi^A_i)+\log(1-\pi^B_i)\right|\right|\right],\]
using the triangle inequality
\[E\left[\left||\log(\pi^B_i)|-|\log(\pi^A_i)|-|\log(1-\pi^B_i)|\right|\right]<E\left[\left|\left|\log(\pi^B_i)\right|-\left|\log(\pi^A_i)+\log(1-\pi^B_i)\right|\right|\right]\]
and eventually (this can be shown numerically)
\[E\left[\left||\log(\pi^B_i)|-|\log(\pi^A_i)|\right|\right]<E\left[\left||\log(\pi^B_i)|-|\log(\pi^A_i)|-|\log(1-\pi^B_i)|\right|\right]<\epsilon_\theta.\]
Next, \(E\left[|\log(\pi^B_i)|-|\log(\pi^A_i)|\right]\le E\left[\left||\log(\pi^B_i)|-|\log(\pi^A_i)|\right|\right]\) and using linearity \(-E\left[|\log(\pi^A_i)|\right]\le \epsilon_\theta - E\left[|\log(\pi^B_i)|\right]\). As probabilities lie in \((0,1)\), we get \[-E\left[|\log(\pi^A_i)|\right]=-E\left[\log\left(\frac{1}{\pi^A_i}\right)\right]=E\left[-\log\left(\frac{1}{\pi^A_i}\right)\right]=E\left[\log(\pi^A_i)\right]\]
so \(E\left[\log(\pi^A_i)\right]\le \epsilon_\theta+E\left[\log(\pi^B_i)\right]\), \(2E\left[\log(\pi^A_i)\right]\le 2\epsilon_\theta+2E\left[\log(\pi^B_i)\right]\) and \(E\left[(\pi^A_i)^2\right]\le \exp(2\epsilon_\theta)E\left[(\pi^B_i)^2\right]\).

Using the symmetry of absolute values we write
\[E\left[\left|\left|\log(1-\pi^B_i)+\log(\pi^A_i)\right|-\left|\log(1-\pi^A_i)+\log(\pi^B_i)\right|\right|\right]<E\left[\left|logit(\pi^B_i)-logit(\pi^A_i)\right|\right],\]
then apply the same steps as before to get \(E\left[(1-\pi^A_i)^2\right]\le \exp(2\epsilon_\theta)E\left[(1-\pi^B_i)^2\right]\). This can also be written as \(E\left[(y_i-\pi^A_i)^2\middle|y_i=1\right]\le \exp(2\epsilon_\theta)E\left[(y_i-\pi^B_i)^2\middle|y_i=1\right]\). Multiplying both sides by \(P(Y=1)\) we get \(E\left[(y_i-\pi^A_i)^2\middle|y_i=1\right]P(Y=1)\le \exp(2\epsilon_\theta)E\left[(y_i-\pi^B_i)^2\middle|y_i=1\right]P(Y=1)\). The bound \(E\left[(\pi^A_i)^2\right]\le \exp(2\epsilon_\theta)E\left[(\pi^B_i)^2\right]\) can be written as \(E\left[(y_i-\pi^A_i)^2\middle|y_i=0\right]P(Y=0)\le \exp(\epsilon_\theta)E\left[(y_i-\pi^B_i)^2\middle|y_i=0\right]P(Y=0)\).

We add up these two bounds to get

\[E\left[(y_i-\pi^A_i)^2\middle|y_i=0\right]P(Y=0)+E\left[(y_i-\pi^A_i)^2\middle|y_i=1\right]P(Y=1)\le\]
\[\exp(2\epsilon_\theta)E\left[(y_i-\pi^B_i)^2\middle|y_i=0\right]P(Y=0)+\exp(2\epsilon_\theta)E\left[(y_i-\pi^B_i)^2\middle|y_i=1\right]P(Y=1)\]
then using Law of Total Expectation:

\[E\left[(y_i-\pi^A_i)^2\right]\le \exp(2\epsilon_\theta)E\left[(y_i-\pi^B_i)^2\right]\]

and finally

\[\frac{E\left[(y_i-\pi^A_i)^2\right]}{E\left[(y_i-\pi^B_i)^2\right]}\le \exp(2\epsilon_\theta).\]

In a similar manner, a lower bound can be obtained and we get that the Brier scores ratio is bounded
\[\exp(-2\epsilon_\theta) \le \frac{E\left[(y_i-\pi^A_i)^2\right]}{E\left[(y_i-\pi^B_i)^2\right]}\le \exp(2\epsilon_\theta),\] meaning performance equivalence is achieved.
\(\blacksquare\)

\hypertarget{sim-stud-res}{%
\subsection{Simulation Study Results}\label{sim-stud-res}}

\begin{figure}[H]

{\centering \includegraphics{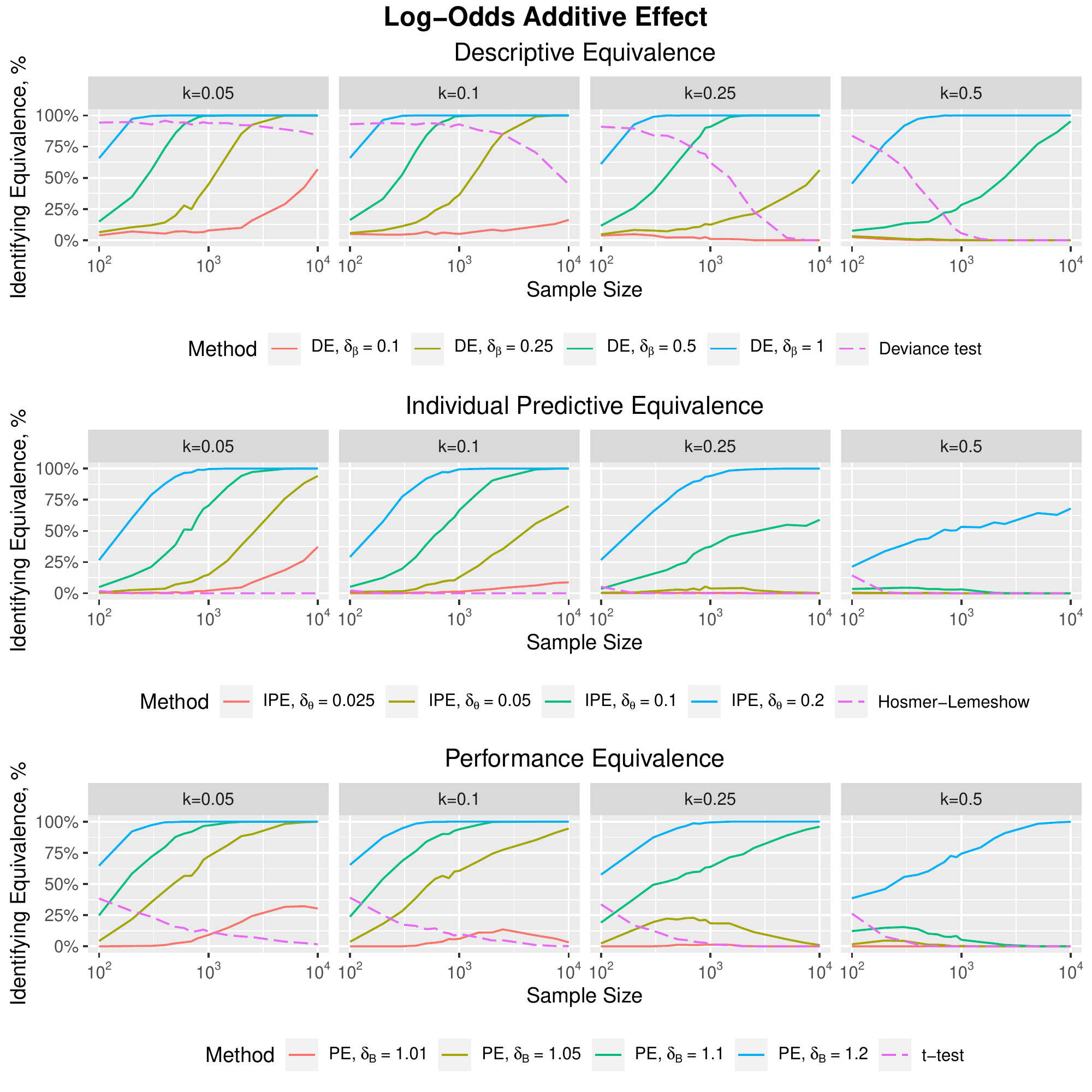} 

}

\caption{Testing the performance of different comparison methods against sample size, under log-odds additive effect. DE stands for descriptive equivalence test (with different $\delta_\beta$ values), IPE stands for individual predictive equivalence test (with different $\delta_\theta$ values), PE stands for performance equivalence test (with different $\delta_B$ values).}\label{fig:sim2-add}
\end{figure}

\begin{figure}[H]

{\centering \includegraphics{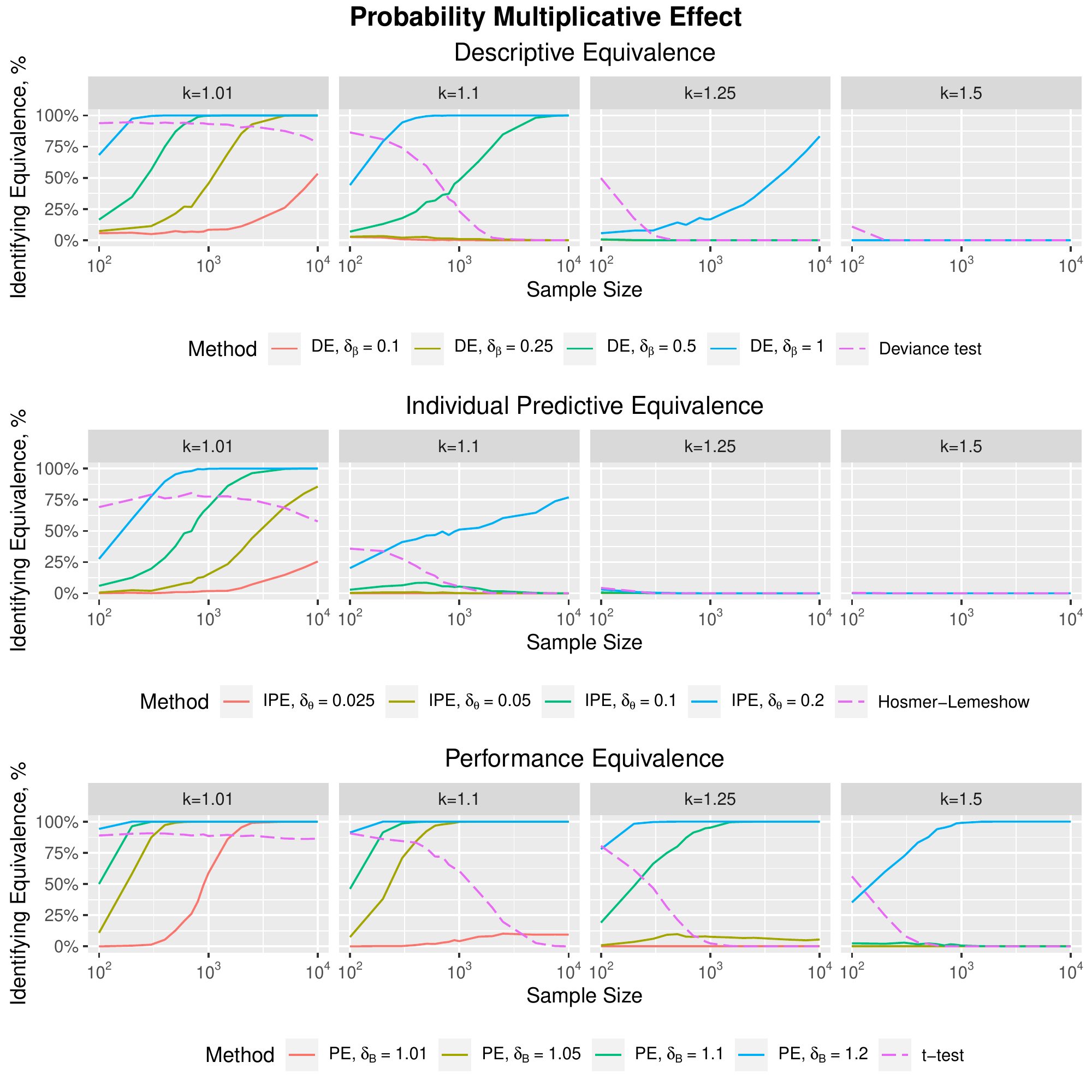} 

}

\caption{Testing the performance of different comparison methods against sample size, under probability multiplicative effect. DE stands for descriptive equivalence test (with different $\delta_\beta$ values), IPE stands for individual predictive equivalence test (with different $\delta_\theta$ values), PE stands for performance equivalence test (with different $\delta_B$ values).}\label{fig:sim3-prob}
\end{figure}

\hypertarget{matal-dg}{%
\subsection{MATAL Data Generation Process}\label{matal-dg}}

The data used to design MATAL comes from two datasets:

\begin{enumerate}
\def\labelenumi{\arabic{enumi}.}
\item
  Norms research dataset -- participants with no learning disabilities, participated in the national norms building for the MATAL tasks.
\item
  Revalidation dataset -- participants who have applied for MATAL-based diagnosis between the years 2008 and 2011.
\end{enumerate}

Each dataset has the following variables: Gender (m/f); clinical diagnosis for dysgraphia (binary); clinical diagnosis for dyscalculia (binary); five exam scores \(x_1,...,x_5\). Each exam score is a linear transformation of a gamma variable: Let \(z_i\sim \Gamma(\alpha_i, \beta_i)\), then \(x_i = C_i - z_i\) with \(C_i\) being some non-negative constant. As the five exam scores are correlated, we should use a multivatiate gamma distribution to generate them. The datasets are of similar size, and are balanced with respecr to gender.

For a given dataset, the following procedure was used for regeneration:

\textbf{Estimation}

\begin{enumerate}
\def\labelenumi{\arabic{enumi}.}
\item
  Invert each variable using \(z_i = max(x_i) - x_i\) (so \(z_i\) is gamma-distributed).
\item
  \(\mu_1,...,\mu_5\) is taken as the vector of means and \(\Sigma=(\sigma_{ij})\) as the covariance matrix.
\item
  Gamma distribution parameters were estimated using the method of moments: \(\hat{\alpha}_i=\frac{\mu^2_i}{{\sigma}^2_{ii}}, \hat{\beta}=\frac{\mu_i}{{\sigma}^2_{ii}}\).
\end{enumerate}

\textbf{Sampling from model}

\begin{enumerate}
\def\labelenumi{\arabic{enumi}.}
\setcounter{enumi}{3}
\item
  Multivariate data matrix \(Y\) was generated using the multivariate normal distribution, with input parameters \(\mu\) and \(\Sigma\).
\item
  Marginal gamma variables \(z^r_{i}\) were created using the gamma distribution quantile function with parameters \(\hat{\alpha}, \hat{\beta}\) and the probabilities vector \(P_i\) of each regenerated variable \(y^r_{i}\).
\item
  Inverted variables were created using \(x^r_i = max(x_i) -z^r_i\)
\end{enumerate}

The Norms dataset was split according to gender, then each subgroup (total of 2 subgroups) was regenerated with \(n=2000\) so overall the regenerated Norms dataset has \(n=4000\) samples. The Revalidation dataset was split according to gender and combination of disabilities (total of 8 subgroups). Each subgroup was regenerated with \(n=500\) so overall the regenerated Revalidation dataset has \(n=4000\) samples.

Regenerated datasets were combined, then split again by gender. Each gender-based dataset (\(n=4000\)) was split to \emph{train} and \emph{test} sets with a 3:1 ratio. The final dataset sizes are \(n_{female}^{train}=n_{male}^{train}=3000, n_{female}^{test}=n_{male}^{test}=1000\).

\newpage

\newpage

\hypertarget{references}{%
\section*{References ~}\label{references}}
\addcontentsline{toc}{section}{References ~}

\hypertarget{refs}{}
\begin{CSLReferences}{1}{0}
\leavevmode\vadjust pre{\hypertarget{ref-R-rmarkdown}{}}%
Allaire, J., Xie, Y., McPherson, J., Luraschi, J., Ushey, K., Atkins, A., et al. (2019). \emph{{rmarkdown}: Dynamic documents for {R}}. \url{https://CRAN.R-project.org/package=rmarkdown}

\leavevmode\vadjust pre{\hypertarget{ref-american2013diagnostic}{}}%
American Psychiatric Association and others. (2013). \emph{Diagnostic and statistical manual of mental disorders ({DSM}-5{\textregistered})}. American Psychiatric Pub.

\leavevmode\vadjust pre{\hypertarget{ref-R-LogRegEquiv}{}}%
Ashiri-Prossner, G. (2022). \emph{{LogRegEquiv}: Logistic regression equivalence}. \url{https://CRAN.R-project.org/package=LogRegEquiv}

\leavevmode\vadjust pre{\hypertarget{ref-R-gridExtra}{}}%
Auguie, B. (2017). \emph{{gridExtra}: Miscellaneous functions for "grid" graphics}. \url{https://CRAN.R-project.org/package=gridExtra}

\leavevmode\vadjust pre{\hypertarget{ref-barker2002assessing}{}}%
Barker, L. E., Luman, E. T., McCauley, M. M., \& Chu, S. Y. (2002). Assessing equivalence: An alternative to the use of difference tests for measuring disparities in vaccination coverage. \emph{American Journal of Epidemiology}, \emph{156}(11), 1056--1061.

\leavevmode\vadjust pre{\hypertarget{ref-benedetti2010scoring}{}}%
Benedetti, R. (2010). Scoring rules for forecast verification. \emph{Monthly Weather Review}, \emph{138}(1), 203--211.

\leavevmode\vadjust pre{\hypertarget{ref-ben2007matal}{}}%
Ben-Simon, A. (2007). \emph{MATAL: A computerized test battery for the diagnosis of learning disabilities}. Jerusalem, Israel: National Institute for Testing \& Evaluation.

\leavevmode\vadjust pre{\hypertarget{ref-ben2013matal}{}}%
Ben-Simon, A. (2013). \emph{MATAL: User guide}. Jerusalem, Israel: National Institute for Testing \& Evaluation.

\leavevmode\vadjust pre{\hypertarget{ref-ben2008regulating}{}}%
Ben-Simon, A., Beyth-Marom, R., Inbar-Weiss, N., \& Cohen, Y. (2008). Regulating the diagnosis of learning disability and the provision of test accommodations in institutions of higher education. In \emph{34th conference of the association for educational assessment cambridge, UK.{[}google scholar{]}}.

\leavevmode\vadjust pre{\hypertarget{ref-berninger2011evidence}{}}%
Berninger, V. W., \& O'Malley May, M. (2011). Evidence-based diagnosis and treatment for specific learning disabilities involving impairments in written and/or oral language. \emph{Journal of Learning Disabilities}, \emph{44}(2), 167--183.

\leavevmode\vadjust pre{\hypertarget{ref-bradley2008sampling}{}}%
Bradley, A. A., Schwartz, S. S., \& Hashino, T. (2008). Sampling uncertainty and confidence intervals for the brier score and brier skill score. \emph{Weather and Forecasting}, \emph{23}(5), 992--1006.

\leavevmode\vadjust pre{\hypertarget{ref-brenner1985evidence}{}}%
Brenner, C. H. (1985). Evidence, probability, and paternity. \emph{American journal of human genetics}, \emph{37}(4), 826.

\leavevmode\vadjust pre{\hypertarget{ref-brier1950verification}{}}%
Brier, G. W. (1950). Verification of forecasts expressed in terms of probability. \emph{Monthly Weather Review}, \emph{78}(1), 1--3.

\leavevmode\vadjust pre{\hypertarget{ref-byrne1988measuring}{}}%
Byrne, B. M. (1988). Measuring adolescent self-concept: Factorial validity and equivalency of the {SDQ III} across gender. \emph{Multivariate Behavioral Research}, \emph{23}(3), 361--375.

\leavevmode\vadjust pre{\hypertarget{ref-byrne1989testing}{}}%
Byrne, B. M., Shavelson, R. J., \& Muthén, B. (1989). Testing for the equivalence of factor covariance and mean structures: The issue of partial measurement invariance. \emph{Psychological bulletin}, \emph{105}(3), 456.

\leavevmode\vadjust pre{\hypertarget{ref-byrne2014factorial}{}}%
Byrne, B. M., \& van de Vijver, F. J. (2014). Factorial structure of the family values scale from a multilevel-multicultural perspective. \emph{International Journal of Testing}, \emph{14}(2), 168--192.

\leavevmode\vadjust pre{\hypertarget{ref-casabianca2018statistical}{}}%
Casabianca, J. M., \& Lewis, C. (2018). Statistical equivalence testing approaches for mantel--haenszel {DIF} analysis. \emph{Journal of Educational and Behavioral Statistics}, \emph{43}(4), 407--439.

\leavevmode\vadjust pre{\hypertarget{ref-chow2008design}{}}%
Chow, S.-C., \& Liu, J. (2008). \emph{Design and analysis of bioavailability and bioequivalence studies}. CRC Press.

\leavevmode\vadjust pre{\hypertarget{ref-collins1998race}{}}%
Collins, J. M., \& Gleaves, D. H. (1998). Race, job applicants, and the five-factor model of personality: Implications for black psychology, industrial/organizational psychology, and the five-factor theory. \emph{Journal of Applied Psychology}, \emph{83}(4), 531.

\leavevmode\vadjust pre{\hypertarget{ref-counsell2015equivalence}{}}%
Counsell, A., \& Cribbie, R. A. (2015). Equivalence tests for comparing correlation and regression coefficients. \emph{British Journal of Mathematical and Statistical Psychology}, \emph{68}(2), 292--309.

\leavevmode\vadjust pre{\hypertarget{ref-cramer2002origins}{}}%
Cramer, J. S. (2002). The origins of logistic regression.

\leavevmode\vadjust pre{\hypertarget{ref-dette2018equivalence}{}}%
Dette, H., Möllenhoff, K., Volgushev, S., \& Bretz, F. (2018). Equivalence of regression curves. \emph{Journal of the American Statistical Association}, \emph{113}(522), 711--729.

\leavevmode\vadjust pre{\hypertarget{ref-devine2013gender}{}}%
Devine, A., Soltész, F., Nobes, A., Goswami, U., \& Szűcs, D. (2013). Gender differences in developmental dyscalculia depend on diagnostic criteria. \emph{Learning and Instruction}, \emph{27}, 31--39.

\leavevmode\vadjust pre{\hypertarget{ref-dolado2014equivalence}{}}%
Dolado, J. J., Otero, M. C., \& Harman, M. (2014). Equivalence hypothesis testing in experimental software engineering. \emph{Software Quality Journal}, \emph{22}(2), 215--238.

\leavevmode\vadjust pre{\hypertarget{ref-greene2008noninferiority}{}}%
Greene, C. J., Morland, L. A., Durkalski, V. L., \& Frueh, B. C. (2008). Noninferiority and equivalence designs: Issues and implications for mental health research. \emph{Journal of traumatic stress}, \emph{21}(5), 433--439.

\leavevmode\vadjust pre{\hypertarget{ref-hauschke1999sample}{}}%
Hauschke, D., Kieser, M., Diletti, E., \& Burke, M. (1999). Sample size determination for proving equivalence based on the ratio of two means for normally distributed data. \emph{Statistics in Medicine}, \emph{18}(1), 93--105.

\leavevmode\vadjust pre{\hypertarget{ref-hauschke2007bioequivalence}{}}%
Hauschke, D., Steinijans, V., \& Pigeot, I. (2007). \emph{Bioequivalence studies in drug development: Methods and applications} (Vol. 60). John Wiley \& Sons.

\leavevmode\vadjust pre{\hypertarget{ref-holland1988differential}{}}%
Holland, P. W., \& Thayer, D. T. (1988). Differential item performance and the mantel-haenszel procedure. \emph{Test validity}, 129--145.

\leavevmode\vadjust pre{\hypertarget{ref-holland2012differential}{}}%
Holland, P. W., \& Wainer, H. (2012). \emph{Differential item functioning}. Routledge.

\leavevmode\vadjust pre{\hypertarget{ref-hosmer2013applied}{}}%
Hosmer Jr, D. W., Lemeshow, S., \& Sturdivant, R. X. (2013). \emph{Applied logistic regression}. John Wiley \& Sons.

\leavevmode\vadjust pre{\hypertarget{ref-jensen1997bounds}{}}%
Jensen, D. (1997). Bounds on mahalanobis norms and their applications. \emph{Linear algebra and its applications}, \emph{264}, 127--139.

\leavevmode\vadjust pre{\hypertarget{ref-joe2014dependence}{}}%
Joe, H. (2014). \emph{Dependence modeling with copulas}. CRC press.

\leavevmode\vadjust pre{\hypertarget{ref-jonkman2009equivalence}{}}%
Jonkman, J. N., \& Sidik, K. (2009). Equivalence testing for parallelism in the four-parameter logistic model. \emph{Journal of biopharmaceutical statistics}, \emph{19}(5), 818--837.

\leavevmode\vadjust pre{\hypertarget{ref-R-ResourceSelection}{}}%
Lele, S. R., Keim, J. L., \& Solymos, P. (2019). \emph{{ResourceSelection}: Resource selection (probability) functions for use- availability data}. \url{https://CRAN.R-project.org/package=ResourceSelection}

\leavevmode\vadjust pre{\hypertarget{ref-liu2010simultaneous}{}}%
Liu, Wei. (2010). \emph{Simultaneous inference in regression}. CRC Press.

\leavevmode\vadjust pre{\hypertarget{ref-liu2009assessing}{}}%
Liu, W., Bretz, F., Hayter, A., \& Wynn, H. (2009). Assessing nonsuperiority, noninferiority, or equivalence when comparing two regression models over a restricted covariate region. \emph{Biometrics}, \emph{65}(4), 1279--1287.

\leavevmode\vadjust pre{\hypertarget{ref-magis2011identification}{}}%
Magis, D., \& De Boeck, P. (2011). Identification of differential item functioning in multiple-group settings: A multivariate outlier detection approach. \emph{Multivariate Behavioral Research}, \emph{46}(5), 733--755.

\leavevmode\vadjust pre{\hypertarget{ref-martinkova2017checking}{}}%
Martinková, P., Drabinová, A., Liaw, Y.-L., Sanders, E. A., McFarland, J. L., \& Price, R. M. (2017). Checking equity: Why differential item functioning analysis should be a routine part of developing conceptual assessments. \emph{CBE---Life Sciences Education}, \emph{16}(2), rm2.

\leavevmode\vadjust pre{\hypertarget{ref-meredith1993measurement}{}}%
Meredith, W. (1993). Measurement invariance, factor analysis and factorial invariance. \emph{Psychometrika}, \emph{58}(4), 525--543.

\leavevmode\vadjust pre{\hypertarget{ref-R-latex2exp}{}}%
Meschiari, S. (2021). \emph{{latex2exp}: Use {LaTeX} expressions in plots}. \url{https://CRAN.R-project.org/package=latex2exp}

\leavevmode\vadjust pre{\hypertarget{ref-ozdemir2015comparison}{}}%
Özdemir, B. (2015). A comparison of IRT-based methods for examining differential item functioning in TIMSS 2011 mathematics subtest. \emph{Procedia-Social and Behavioral Sciences}, \emph{174}, 2075--2083.

\leavevmode\vadjust pre{\hypertarget{ref-peng2002introduction}{}}%
Peng, C.-Y. J., Lee, K. L., \& Ingersoll, G. M. (2002). An introduction to logistic regression analysis and reporting. \emph{The journal of educational research}, \emph{96}(1), 3--14.

\leavevmode\vadjust pre{\hypertarget{ref-putnick2016measurement}{}}%
Putnick, D. L., \& Bornstein, M. H. (2016). Measurement invariance conventions and reporting: The state of the art and future directions for psychological research. \emph{Developmental Review}, \emph{41}, 71--90.

\leavevmode\vadjust pre{\hypertarget{ref-robinson2005regression}{}}%
Robinson, A. P., Duursma, R. A., \& Marshall, J. D. (2005). A regression-based equivalence test for model validation: Shifting the burden of proof. \emph{Tree physiology}, \emph{25}(7), 903--913.

\leavevmode\vadjust pre{\hypertarget{ref-siqueira2008active}{}}%
Siqueira, A. L., Whitehead, A., \& Todd, S. (2008). Active-control trials with binary data: A comparison of methods for testing superiority or non-inferiority using the odds ratio. \emph{Statistics in medicine}, \emph{27}(3), 353--370.

\leavevmode\vadjust pre{\hypertarget{ref-steinberg2006using}{}}%
Steinberg, L., \& Thissen, D. (2006). Using effect sizes for research reporting: Examples using item response theory to analyze differential item functioning. \emph{Psychological methods}, \emph{11}(4), 402.

\leavevmode\vadjust pre{\hypertarget{ref-stevens2017comparing}{}}%
Stevens, N. T., \& Anderson-Cook, C. M. (2017a). Comparing the reliability of related populations with the probability of agreement. \emph{Technometrics}, \emph{59}(3), 371--380.

\leavevmode\vadjust pre{\hypertarget{ref-stevens2017quantifying}{}}%
Stevens, N. T., \& Anderson-Cook, C. M. (2017b). Quantifying similarity in reliability surfaces using the probability of agreement. \emph{Quality Engineering}, \emph{29}(3), 395--408.

\leavevmode\vadjust pre{\hypertarget{ref-swaminathan1990detecting}{}}%
Swaminathan, H., \& Rogers, H. J. (1990). Detecting differential item functioning using logistic regression procedures. \emph{Journal of Educational measurement}, \emph{27}(4), 361--370.

\leavevmode\vadjust pre{\hypertarget{ref-van2015measurement}{}}%
van de Schoot, R., Schmidt, P., De Beuckelaer, A., Lek, K., \& Zondervan-Zwijnenburg, M. (2015). Measurement invariance. \emph{Frontiers in psychology}, \emph{6}, 1064.

\leavevmode\vadjust pre{\hypertarget{ref-vandenberg2000review}{}}%
Vandenberg, R. J., \& Lance, C. E. (2000). A review and synthesis of the measurement invariance literature: Suggestions, practices, and recommendations for organizational research. \emph{Organizational research methods}, \emph{3}(1), 4--70.

\leavevmode\vadjust pre{\hypertarget{ref-verhagen2013bayesian}{}}%
Verhagen, A., \& Fox, J. (2013). Bayesian tests of measurement invariance. \emph{British Journal of Mathematical and Statistical Psychology}, \emph{66}(3), 383--401.

\leavevmode\vadjust pre{\hypertarget{ref-walker2011understanding}{}}%
Walker, E., \& Nowacki, A. S. (2011). Understanding equivalence and noninferiority testing. \emph{Journal of general internal medicine}, \emph{26}(2), 192--196.

\leavevmode\vadjust pre{\hypertarget{ref-weigold2016equivalence}{}}%
Weigold, A., Weigold, I. K., Drakeford, N. M., Dykema, S. A., \& Smith, C. A. (2016). Equivalence of paper-and-pencil and computerized self-report surveys in older adults. \emph{Computers in Human Behavior}, \emph{54}, 407--413.

\leavevmode\vadjust pre{\hypertarget{ref-wellek2010testing}{}}%
Wellek, S. (2010). \emph{Testing statistical hypotheses of equivalence and noninferiority}. Chapman; Hall/CRC.

\leavevmode\vadjust pre{\hypertarget{ref-wells2009range}{}}%
Wells, C. S., Cohen, A. S., \& Patton, J. (2009). A range-null hypothesis approach for testing DIF under the rasch model. \emph{International Journal of Testing}, \emph{9}(4), 310--332.

\leavevmode\vadjust pre{\hypertarget{ref-R-ggplot2}{}}%
Wickham, H., Chang, W., Henry, L., Pedersen, T. L., Takahashi, K., Wilke, C., et al. (2019). \emph{{ggplot2}: Create elegant data visualisations using the grammar of graphics}. \url{https://CRAN.R-project.org/package=ggplot2}

\leavevmode\vadjust pre{\hypertarget{ref-R-bookdown}{}}%
Xie, Y. (2019b). \emph{{bookdown}: Authoring books and technical documents with {R} markdown}. \url{https://github.com/rstudio/bookdown}

\leavevmode\vadjust pre{\hypertarget{ref-R-knitr}{}}%
Xie, Y. (2019a). \emph{{knitr}: A general-purpose package for dynamic report generation in {R}}. \url{https://CRAN.R-project.org/package=knitr}

\leavevmode\vadjust pre{\hypertarget{ref-R-kableExtra}{}}%
Zhu, H. (2019). \emph{{kableExtra}: Construct complex table with 'kable' and pipe syntax}. \url{https://CRAN.R-project.org/package=kableExtra}

\end{CSLReferences}

\end{document}